\documentclass[12pt,a4paper]{article}
\pdfoutput=1
\usepackage{amsmath}
\usepackage{graphicx}
\usepackage{subfigure}
\usepackage{bm}
\usepackage{tipa}
\usepackage{url}
\usepackage{natbib}
\usepackage{lineno}
\usepackage{geometry}
\usepackage{bbold}

\newcommand{\imp}[1]{\mathcal{I}(#1)}
\newcommand{\impf}[2]{\mathcal{I}_{#1}(#2)}
\newcommand{\var}[1]{\mathrm{Var}\left[#1\right]}

\newcommand{\E}[1]{\mathrm{E}\left[#1\right]}

\newcommand{\tspace}{\mathcal{X}}
\newcommand{\tspacef}[1]{\mathcal{X}_{#1}}
\newcommand{\tsubspace}[1]{\mathcal{X}^{(#1)}}
\newcommand{\tsubspacef}[2]{\mathcal{X}^{(#1)}_{#2}}
\newcommand{\nruns}{m}
\newcommand{\nchains}{n}
\newcommand{\ilevel}{b}
\newcommand{\dens}[2]{\pi_{#1}(#2)}
\newcommand{\densno}[1]{\pi(#1)}
\newcommand{\xvec}{\bm{x}}
\newcommand{\yvec}{\bm{y}}
\newcommand{\vrw}[2]{V_{#1,#2}}
\newcommand{\whole}{\mathrm{whole}}
\newcommand{\indicator}{\mathbb{1}}
\newcommand{\Number}[1]{Z(#1)}
\newcommand{\samples}[2]{X^{#1}_{#2}}
\newcommand{\nsamples}{s}
\newcommand{\pvol}{p}
\newcommand{\hpvol}{\hat{\pvol}}

\newcommand{\be}{\begin{equation}}
\newcommand{\ee}{\end{equation}}
\newcommand{\ba}{\begin{eqnarray}}
\newcommand{\ea}{\end{eqnarray}}
\newcommand{\bi}{\begin{itemize}}
\newcommand{\ei}{\end{itemize}}
\newcommand{\bfi}{\begin{figure}}
\newcommand{\efi}{\end{figure}}
\newcommand{\bn}{\begin{enumerate}}
\newcommand{\en}{\end{enumerate}}

\def\acool{\alpha_{\rm cool}}

\def\arh{\alpha_{\rm reheat}}
\def\alphahot{\alpha_{\rm hot}}

\def\alphareheat{{\arh}}
\def\alphacool{{\acool}}

\def\VCUT{v_{\rm cut}}
\def\ZCUT{z_{\rm cut}}

\def\fellip{f_{\rm ellip}}
\def\fburst{f_{\rm burst}}

\def\yield{p_{\rm yield}}
\def\stabledisk{f_{\rm stab}}
\def\epsilonSMBHEddington{\epsilon_{\rm Edd}}

\title{Efficient uniform designs for multi-wave computer experiments}
\author{Daniel Williamson$^{1}$, Ian Vernon$^{2}$ \\ $^{1}$ College of Engineering, Mathematics and Physical Sciences, \\ University of Exeter, Exeter, UK \\ $^{2}$ Department of Mathematical Sciences, \\ Durham University, Durham, UK}

\begin{document}
\maketitle
\begin{abstract}
In this paper we tackle the problem of generating uniform designs in very small subregions of computer model input space that have been identified in previous experiments as worthy of further study. The method is capable of producing uniform designs in subregions of computer model input space defined by a membership function that consists of a continuous function passing a threshold test, and does so far more efficiently than current methods when these subregions are small. Our application is designing for regions of input space that are not ruled out by history matching, a statistical methodology applied in numerous diverse scientific applications whereby model runs are used to cut out regions of input space that are incompatible with real world observations. History matching defines a membership function for a region of input space that is not ruled out yet by observations in the form of a distance metric called implausibility. We use this distance metric to drive a new type of Evolutionary Monte Carlo algorithm with a uniform distribution on the not ruled out yet region as its target distribution. The algorithm can locate and generate uniform points within extremely small subspaces of the computer model input space with complex and even disconnected topologies. We illustrate the performance of the technique in comparison to current methods with a number of idealised examples. We then apply our algorithm to generating an optimal design for the not ruled out yet region of a galaxy simulation model called GALFORM following 4 previous waves of history matching where the target region is 0.001\% the volume of the input space.
\end{abstract}


\section{Introduction}
There is a large literature on generating designs for computer experiments when the input space, assumed to have dimension $d$, can be scaled to the d-dimensional unit hypercube $[0,1]^{d}$ by way of a linear transformation. These methods are useful for obtaining an initial set of computer model runs. Current thinking on initial designs for computer experiments centres around the idea of generating a design that is ``space filling'', in the sense that, given the number of model runs available, a design is chosen so that as much of the input space is explored as possible. The current method of choice is based on Latin Hypercube (LH) sampling whereby, for an $n-$point design, the range of each input is divided into $n$ intervals according to some distribution, and where each interval is represented exactly once in the resulting design \citep[see, for example][]{santneretal03}. \par LH sampling defines a class of ``space filling'' designs from which a design is usually generated according to some other criterion. The most popular of these is a design based on finding the LH that maximises the minimum distance between design points called the Maximin Latin Hypercube \citep{morrismitchell95}, though others exist, based on harmonic means \citep{zaglauer12} and on orthogonality of the resulting design \citep{linetal09}. \par  

Whilst Latin Hypercubes may be good initial designs for exploring input spaces, it will rarely be the case that the best possible design for a computer experiment would be based on a single Latin Hypercube. Depending on the goal of the experiment, it will almost always be of benefit to divide the overall budget of model runs into waves of runs sequentially designed to satisfy the experimental goals. A considerable amount of research has taken place in the optimisation literature, for example, based on constructing sequential designs that seek to find the minimum of some function of the computer model output. Much of this is based on the idea of expected improvement, where, following an initial space filling design, subsequent waves are designed to maximise a criterion based on the predicted reduction of the minimum with respect to the remaining uncertainty in the function \citep{jonesetal98, santneretal03}. \par 

The focus of this paper is on a technique to help with sequential designs for computer experiments where the goal is to identify regions of the input space that represent computer model output that is not inconsistent with real world observations or with desired model behaviour. In these experiments, following an initial space filling design, a membership rule can be developed identifying a region of interest within the initial input space. Given this rule, we could determine whether or not a setting of the model inputs lies within this region. Though there may be many ways to develop such membership rules, we will focus on one that is gaining popularity within the computer experiment community called history matching.

\par History matching (described in detail in section \ref{history.matching}), is a method that uses emulators (fast statistical representations of computer models that include a measure of uncertainty in the prediction of the computer model output), to compare real world observations and the computer model output using a measure based on the standardised distance between the observations and the expected output \citep{craigetal96}. The measure is used to define a membership rule where the space that can be ruled out has a value of the measure larger than a defined threshold. History matching therefore defines subspaces of the computer model input space that are of interest because they cannot be ruled out using the current model data and observations. History matching has been applied to the simulation of oil wells \citep{craigetal97}, to the simulation of galaxy formation \citep{vernonetal10} and to climate models \citep{edwardsetal11, williamsonetal13}. \par

Successful application of history matching means iteratively cutting down the model input space in waves. At each wave a design is created within the subspace defined by the previous wave and is used to history match again and further reduce the volume of the not ruled out yet space. As yet, the design of these subsequent waves represents an open and challenging problem. Currently, only rejection sampling is used to generate designs for subsequent waves. \cite{vernonetal10} describe a method of design that involves generating latin hypercubes over the full space and only keeping those points that are not ruled out by history matching. This process is repeated until the desired number of design points is obtained. As the target subspace is only defined by a membership rule, there is no way of determining what properties, if any, designs of this nature have. They may, or may not, offer a good coverage of the input space and there is no way to tell. \par

Furthermore, designs of this nature can be extremely expensive (or even impossible) to generate when the volume of the target subspace is small relative to the initial space. Though a feature of history matching (and indeed any method based on emulators) is that the membership function is cheap relative to running the computer model, even a function that takes up to a few seconds could make it very unlikely to hit extremely small regions enough times to generate a large enough design. The application we present in this paper has a subspace with a volume of order $10^{-5}$ of the original space, though we have dealt with spaces with many orders of magntiude smaller relative volumes in other applications. 

\par In this paper we develop an algorithm for efficient generation of uniform random samples from small subspaces of a computer model input space defined by a membership function. We present this in the context of history matching, but note its applicability to any method containing a membership rule based on a continuous function passing a threshold test. We combine the ideas of Evolutionary Monte Carlo \citep{liangwong01}, a method from the MCMC literature, with our membership rule to develop an algorithm which we call \textit{implausibility driven evolutionary Monte Carlo}. We illustrate our procedure using 3 idealised examples and then apply the algorithm to obtaining a sample from a very small subspace of a galaxy simulation model called GALFORM.
\par In section \ref{history.matching} we describe history matching and implausibility. In section \ref{idemc.description} we adapt evolutionary Monte Carlo to develop our implausibility driven evolutionary Monte Carlo algorithm. Section \ref{ladder.choice} discusses issues with initialisation of the algorithm and presents a method for defining an initial population that leads to an efficient sampler. In section \ref{examples} we apply our method in 3 illustrative examples. In section \ref{multi.wave} we extend the algorithm to the case where there are multiple waves of history matching and compare its efficiency to rejection sampling. In section \ref{application} we apply our method to generating a uniform design from a very small subspace of the galaxy simulation model GALFORM following 4 previous waves of history matching. Section \ref{discussion} contains discussion and there is an appendix showing mixing plots for some of our examples and our application.

\section{History Matching}\label{history.matching}
We write our computer model as the vector-valued function, $f(x)$, where $x$ is a $d$-dimensional vector of model inputs occupying input space $\tspace$. We assume that this space is defined by specifying a range of values that each input, $x_{i}$, $i=1,\ldots,d,$ is allowed to take so that we may assume that $\tspace$ can be scaled to the hypercube $[0,1]^{d}$. \par
The computer model is designed to represent a complex physical system, $y$, for which we have imperfect observations, $z$. History matching is an established statistical method for ruling out regions of $\tspace$, $\tspace_{RO}$, where $f(x_{0})$, for $x_{0}\in\tspace_{RO}$ is inconsistent with $z$ \citep{craigetal96}. In order to do this, a statistical model relating $f(x)$, $y$ and $z$ is required. A popular choice of statistical model is the `best input' model \citep{kennedyohagan01}, which specifies that there exists $x^{*} \in \tspace$ such that
\begin{equation}\label{best.input}
y = f(x^{*}) + \eta
\end{equation}
where the model discrepancy term, $\eta$, is independent from $x^{*}$ and with $f(x)$ for all $x \in \tspace$. An alternative model linking reality to a computer model is presented in \cite{williamsonetal13}. Detailed discussion of the best input model and possible alternatives is presented in \cite{goldsteinrougier09}. The observations are usually related to the physical system via
\begin{equation}\label{observations}
z = y  + e,
\end{equation}
where $e$ is independent observation error. \par
The computer models we are interested in applying history matching to are either computationally expensive, have high dimensional input spaces or both. This means that we cannot evaluate $f(x)$ at all of the points we need to. We therefore use a computer experiment to build a statistical model for $f(x)$ called an emulator. \par

An emulator is a stochastic representation of a computer model that gives a prediction of the value $f(x)$ with an associated uncertainty for any $x\in \tspace$. The prediction takes a fraction of the time that would be required to evaluate $f(x)$. Typically, an emulator for output $i$ of a computer model has the form
\begin{equation}\label{emulator}
f_{i}(x) = \sum_{i}\beta_{ij}g_{j}(x) + \epsilon_{i}(x)
\end{equation}
where $g(x)$ is a vector of specified functions of $x$, $\beta$ is a matrix of uncertain coefficients, and $\epsilon(x)$ is a mean-zero Gaussian process with a specified form of covariance function. Emulation proceeds by choosing the functions inside $g(x)$ and providing a joint uncertainty specification for $\{\beta, \epsilon(x)\}$. This is achieved using an $\nruns$-point computer experiment $f(x_{1}),\ldots,f(x_{\nruns})$ at design points $x_{j} \in \tspace, j=1,\ldots,\nruns$, and various model fitting techniques \citep[see, for example,][]{sacksetal89, rougier08, williamsonetal12, santneretal03}. \par

History matching rules out regions of $\tspace$ using a membership function. For any $x_{0}\in\tspace$ we can decide whether or not $x_{0}$ is a good candidate for $x^{*}$ by examining the distance between $z$ and our expectation for $f(x_{0})$. The distance we use is called an implausibility measure, $\imp{x_{0}}$, and is either defined on each scalar component of $f(x_{0})$ via
\begin{equation}\label{implausibility.scalar}
\impf{i}{x_{0}} = \frac{|z_{i}-\E{f_{i}(x_{0})}|}{\sqrt{\var{z_{i} - \E{f_{i}(x_{0})}}}},
\end{equation}
with overall implausibility for $f(x_{0})$ given as, for example, $\imp{x_{0}}=\mathrm{max}_{i}\{\impf{i}{x_{0}}\}$; or it is defined by the multivariate analogue
\begin{equation}\label{implausibility.multi}
\imp{x_{0}} = (z - \E{f(x_{0})})^{T}\var{z - \E{f(x_{0})}}^{-1}(z - \E{f(x_{0})}).
\end{equation}
The choice between these two definitions depends on the level of detail of our statistical modelling and, in particular, on our ability and willingness to specify a full covariance structure over each vector required for (\ref{implausibility.multi}). \par 

Large values of $\imp{x_{0}}$ suggest that it is implausbile that $x_{0}$ is consistent with our uncertainty specification and with the observations $z$, so that $x_{0}$ can be ruled out as a candidate for $x^{*}$. We make this formal by choosing a threshold, $a$, so that any $x \in \tspace$ with a value of $\imp{x}>a$ is ruled out. The remaining space we term Not Ruled Out Yet (NROY) space, where NROY space is defined as
\begin{displaymath}
\tspacef{NROY} =  \{x\in\tspace : |\imp{x}|\leq a\}.
\end{displaymath}
The choice of $a$ is problem dependent and also depends on which definition of $\imp{x}$ has been  used. In previous studies when using definition (\ref{implausibility.scalar}) we have set $a=3$ \citep{craigetal97, williamsonetal13}, justifying our choice using Pukelsheim's 3 sigma rule \citep{pukelsheim94}. \par When done iteratively through successive ``waves'' of designs, history matching is at its most powerful. Following an initial history match, a new $\nruns_{1}$-point design in $\tspacef{NROY}$ is constructed and the runs used to build more informative emulators for the same outputs as in the first match, perhaps including additional outputs for which we have corresponding real world observations. These new wave 2 emulators can be used to history match again, reducing NROY space further. This iterative process is termed refocussing. \par
\cite{vernonetal10} history matched a computer model simulating the evolution of galaxies since the big bang using $5$ waves, and explicitly demonstrated the benefits of using such an iterative strategy. The approach is powerful because in each subsequent wave the emulators are expected to improve in accuracy for the following three reasons:
\begin{enumerate}
\item In many computer model applications the function $f(x)$ is assumed to be smooth. At each wave we zoom into a smaller part of this function and hence would expect it to appear smoother still. Taylor series arguments ensure that we would therefore expect $f(x)$ to be well approximated by the linear model part of the emulator (equation~(\ref{emulator})), which is often constructed from low order polynomials in $x$. 
\item There is a higher density of runs in the new NROY space as it is of smaller volume than in the previous wave. This results in the Gaussian process part of the emulator having greater accuracy and hence being able to describe more detailed structure.
\item Emulator construction often involves the identification of active inputs, with the inactive inputs demoted to contributing an uncorrelated noise term. In early waves it is often difficult to identify more that a small set of dominant active variables. However, in later waves these inputs have had their effects curtailed, and hence we can also identify new active inputs with more subtle effects. 
\end{enumerate}
Due to this progressive increase in accuracy of the emulators (and the possible inclusion of more outputs), we expect the NROY space to shrink in each wave.
This process leads to a nested set of emulators, with the wave 1 emulators being of moderate accuracy but applicable to the whole input space $\tspace$, while the later wave emulators, being highly accurate but only applicable to small volumes of the input space, centred around regions where good matches are suspected to be found (i.e. $\tspacef{NROY}$, defined for the appropriate wave). 

These iterative history matching techniques have been successfully applied to several diverse computer models across a range of scientific disciplines including 
galaxy formation simulations (for an overview see~\cite{StatSci13} or see \cite{bower10} for the scientific case), 
oil reservoir simulations (e.g. \cite{craigetal97}, \cite{cumminggoldstein10}), systems biology models (\cite{Stoch10},\cite{Stoch13}), climate models~\citep{williamsonetal13} and rainfall runoff models~\citep{asses_MD}.
Although successful, this approach has a fundamental problem as we now describe, the solution of which being the core motivation for this paper.

Each wave in the refocussing process defines an implausibility function $\impf{[j]}{x}$ using either definition (\ref{implausibility.scalar}) or (\ref{implausibility.multi}) and the emulator built at that wave. NROY space is then defined as 
\begin{equation}\label{NROY.multi}
\tspacef{NROY} = \{x\in\tspace: |\impf{[k]}{x}| \leq a_{k}, \forall k \in 1,\ldots,j\}.
\end{equation} 

After multiple waves of matching, the resulting NROY space can be a tiny fraction of $\tspace$. As we only have a membership function, constructing a well designed set of new points in NROY space, either for further history matching, Bayesian calibration \citep{kennedyohagan01}, or simply to run the model in NROY space to see how it behaves, can be difficult. \par Currently, the way this is done is via rejection sampling. In \cite{vernonetal10}, large Latin Hypercubes are used to sample points from $\tspace$, and those in NROY space are retained. In their study $1000$ points were used to construct emulators at each wave. At each wave, new Latin Hypercubes were proposed until $1000$ unique points in NROY space were found. Latin Hypercubes were used as it was thought that this would ensure the resulting design was well spread in NROY space, though there is no guarantee of this. \par

It is possible to obtain a uniform sample from NROY space using standard Monte Carlo sampling from $\tspace$. A random point is chosen from $\tspace$ and, if it is in NROY space, it is retained. Though this method guarantees uniformity of the sample, it may be extremely difficult to obtain large samples in tiny subspaces, even with emulators, due to low acceptance probabilities. It is also extremely inefficient in that information from previous emulator evaluations that is useful for the search for NROY space is discarded without being used. \par We present an alternative algorithm based on the ideas of evolutionary Monte Carlo, that samples a uniform set of points from NROY space. Our algorithm is particularly efficient for small subspaces of $\tspace$ and offers an estimate of the relative volume of NROY space.

\section{Implausibility Driven Evolutionary Monte Carlo}\label{idemc.description}
We first state our algorithm for the case where we have one implausibility, $\imp{x}$, and where our target distribution is uniform on the set $\{x \in \tspace: |\imp{x}|\leq a\}$. We describe a natural generalisation of the algorithm to implausibility from multiple waves defined as in (\ref{NROY.multi}) in section~\ref{multi.wave}. \par The concept of the algorithm is to adapt the ideas of temperature based MCMC algorithms such as parallel tempering \citep[PT,][]{geyer91} and evolutionary Monte Carlo \citep[EMC,][]{liangwong01}, where parallel chains sample from ``heated'' versions of a target density controlled by a temperature ladder, to implausibility levels. In PT and EMC algorithms a population of densities 
\begin{displaymath}
h_{i}(x) \propto \exp\{-H(x)/t_{i}\}
\end{displaymath}
for some function $H(\cdot)$ and temperatures $t_{i}$ is constructed for $i = 1,...,\nchains$. The temperatures and $H(\cdot)$ are chosen so that the target density is $h_{\nchains}(x)$ for $t_{\nchains}=1$ and so that higher temperatures essentially ``flatten'' the target, making it easier for an MCMC at heated versions of the target density to make large jumps in the state space. These algorithms then run $n$ parallel MCMCs each sampling from $h_{i}(\cdot)$ for $i=1,\ldots,\nchains,$ and allow ``swapping moves'' for adjacent chains every few steps. We formalise this in the context of our algorithm below.
\par To adapt the ideas of PT and EMC to finding uniform NROY designs, we exploit a key property of implausibility in that, for any real $\ilevel_{1}$ and $\ilevel_{2}$ with $\ilevel_{2} \geq \ilevel_{1}$,
\begin{displaymath}
\{x \in \tspace: |\imp{x}|\leq \ilevel_{1}\} \subseteq \{x \in \tspace: |\imp{x}|\leq \ilevel_{2}\}.
\end{displaymath}
By choosing an appropriate implausibility ladder $a=\ilevel_{n}\leq\cdots\leq \ilevel_{1}$, we can set up $\nchains$ parallel chains each sampling from the region defined by the different implausibility cutoff points $\ilevel_{1},\ldots,\ilevel_{\nchains}$. \par 

Suppose we have selected $\ilevel_{1},\ldots,\ilevel_{\nchains}=a$. Let 
\begin{equation}\label{defn.subspaces}
\tsubspace{i} = \{x \in \tspace: |\imp{x}|\leq \ilevel_{i}\}
\end{equation}
and define a sequence of densities
\begin{displaymath}
\dens{i}{x} \sim \mathrm{Uniform}(x\in\tsubspace{i}),
\end{displaymath}
so that sampling from $\dens{\nchains}{\cdot}$ represents a uniform sample from NROY space. The target density of the Markov chain defined on the $\nchains$-fold cartesian product of the original input space is 
\begin{displaymath}
\densno{\xvec} = \prod_{i=1}^{\nchains}\dens{i}{x_{i}}, \qquad \xvec= (x_{1}, \ldots, x_{n})
\end{displaymath}
and each $x_{j}\in\tsubspace{j}$ is a member of the population $\xvec$ called a chromosome. \par EMC evolves the Markov chain $\xvec$ using three types of update move: mutation, crossover and exchange, each described below. Each move preserves the stationary distribution of the Markov chain as long as the move is accepted with probability given by the appropriate Metropolis-Hastings ratio for the proposed move. We describe each move below in the context of a general EMC algorithm and derive the appropriate acceptance probability for our implausibility driven application. More detail on EMC algorithms can be found in \cite{liangwong01}, \cite{goswamiliu07}, and \cite{liangetal10}.
\subsection{Mutation}
In a mutation move, a chromosome, $x_{i}$ is updated via Metropolis Hastings with transition probability $q_{i}(\cdot |x_{i})$ and the proposed value, $y_{i}$, is accepted with probability
\begin{equation}\label{mutate.rate}
r_{M} = \mathrm{min}\left(\frac{\dens{i}{y_{i}}q_{i}(x_{i}|y_{i})}{\dens{i}{x_{i}}q_{i}(y_{i}|x_{i})},1\right).
\end{equation}
A desirable feature in our application of EMC moves is that the acceptance rate is $1$ or very close to $1$ if $y_{i}\in\tsubspace{i}$. This ensures that the information from new, previously unseen points in the target spaces is almost always used to guide us towards NROY space. Because $x_{i}\in\tsubspace{i}$, we can write $r_{M}$ as 
\begin{displaymath}
r_{M} =
\begin{cases}
\mathrm{min}\left(\frac{q_{i}(x_{i}|y_{i})}{q_{i}(y_{i}|x_{i})},1\right) & \text{if } y_{i} \in \tsubspace{i} \\
0 & \text{otherwise}
\end{cases}
\end{displaymath}
Symmetric proposals, with $q_{i}(y_{i}|x_{i}) = q_{i}(x_{i}|y_{i})$, are therefore good choices here, as if a proposed $y_{i}$ is in $\tsubspace{i}$ it will always be accepted. This means that we can control the acceptance rate of mutation moves by tuning the proposal, with a good proposal being such that the acceptance rate is somewhere between $0.2$ and $0.5$ \citep[see][]{gelmanetal96}. For this reason we choose random walk proposals, however, we don't use a simple random walk proposal because the shape of the target space $\tsubspace{i}$ may be complicated and/or disconnected and hence poorly represented by a normal random walk $y_{i}\sim\mathrm{N}(x_{i},V_{i})$ for some variance matrix $V_{i}$. Our solution is to have a proposal that is almost symmetric but one that is closer to $\tsubspace{i}$ than a standard random walk proposal. \par For each $i=1,\ldots,\nchains$ we construct a finite partition $\tspace = \tsubspacef{i}{1}\cup\cdots\cup\tsubspacef{i}{r_{i}}$, and let the variance of a random walk proposal depend on which part of $\tsubspace{i}$ you start in. Let the variance of a random walk move from $x_{i}$ be $\vrw{i}{j}$ for $x_{i}\in\tsubspacef{i}{j}$ and $j=1,\ldots,r_{i}$. Let the variance of a random walk in $\tsubspace{i}$ be $\vrw{i}{\whole}$, then we let our proposal distribution be the mixture
\begin{displaymath}
y_{i}|x_{i} \sim \omega^{i}\mathrm{N}(x_{i},\sum_{j=1}^{r_{i}}\indicator_{x_{i}\in\tsubspacef{i}{j}}\vrw{i}{j}) + (1-\omega^{i})\mathrm{N}\left(x_{i},\vrw{i}{\whole}\right)
\end{displaymath}
with user chosen weight $\omega^{i}$ and where $\indicator_{A}$ is an indicator function for event $A$. \par This proposal ensures that if $\omega^{i}$ is large (say $> 0.8$), then most of the time for $x_{i}\in\tsubspacef{i}{j}$, we propose $y_{i}\in\tsubspacef{i}{j}$. This also implies that when $y_{i}\in\tsubspacef{i}{j}$ after $x_{i}\in\tsubspacef{i}{j}$, $r_{M}=1$ if $y_{i}\in\tsubspace{i}$. The global move, chosen with probability $1-\omega^{i}$, allows $y_{i}$ to more easily jump to different parts of the partition, though the acceptance rate for this type of move may be very small if $\tsubspace{i}$ has a strange shape or is disconnected, as can often happen in history matching.
\par
If $x_{i}\in\tsubspacef{i}{j}$ and $y_{i}\in\tsubspacef{i}{k}$ with $k\neq j$ and if $y_{i}\in\tsubspace{i}$ then
\begin{displaymath}
r_{M} = \mathrm{min}\left(\frac{\omega^{i}g(x_{i};y_{i}, \vrw{i}{k})+(1-\omega^{i})g(x_{i};y_{i},\vrw{i}{\whole})}{\omega^{i}g(y_{i};x_{i}, \vrw{i}{j})+(1-\omega^{i})g(y_{i};x_{i},\vrw{i}{\whole})},1\right)
\end{displaymath}
with $g(\cdot;\mu,\sigma)$ being the density of $N(\mu,\sigma)$. In this situation both $g(x_{i};y_{i},\vrw{i}{k})$ and $g(y_{i};x_{i},\vrw{i}{j})$ will be very small relative to $g(y_{i};x_{i},\vrw{i}{\whole})=g(x_{i};y_{i},\vrw{i}{\whole})$, because $x_{i}$ and $y_{i}$ are in different parts of the partition of $\tspace$. Both the numerator and the denominator of $r_{M}$ will therefore be close to $(1-\omega^{i})g(y_{i};x_{i},\vrw{i}{\whole})$ and so $r_{M}$ will be close to $1$. 
\par 
In high dimensional applications, where many of the inputs have little affect on the output, we have found that there can be an issue using a mixture of normal proposals due to lack of restriction on the support of the proposals leading to a large number of proposed moves outside of $\tspace$. In these cases we have had success using truncated normal distributions with the support defined within the boundaries of $\tspace$ only. This comes with a computational cost (it is more expensive to sample from and evaluate the distribution function of multivariate truncated normal distributions) and at the price of having $q_{i}(y_{i}|x_{i}) \neq q_{i}(x_{i}|y_{i})$ so that we may reject NROY points more often. The code we have provided comes with the option to use truncated normals, though this option was not required for any of the examples or the application in this paper.

\subsection{Real Crossover}
A crossover operation aims to swap individual elements of two randomly chosen chromosomes in order that new and potentially interesting parts of each $\tsubspace{i}$ may be explored. The two chosen chromosomes are referred to as \textit{parent} chromosomes and the samples generated by a crossover operation are called \textit{child} chromosomes. \par Crossover operations may be particularly useful when history matching leads to connected but strangely shaped subspaces of $\tspace$ and when mutation moves that attempt to walk in all $d$ dimensions are either difficult to make, or easy to make in some directions, but difficult to make in others. For example, there may be an offshoot of the main part of NROY space that mainly varies in fewer dimensions than the main part. Mutation moves will be unlikely to explore these parts of the space, but crossover moves may stand a much better chance. \par The crossover move we describe here is the simplest crossover move, called real crossover. Other more complex moves, such as the snooker move \citep{liangwong01} may also be adaptable to the search for NROY space. The first step is to randomly choose two chromosomes to cross, by selecting $i,j\in \{1,\ldots,\nchains\}$, $i\neq j$, with probabilities $p(I=i|\xvec)$ and $p(J=j|\xvec,I=i)$. In one point crossover, we then choose a crossover point, $c\in\{1,\ldots,d\}$, and then define the child chromosomes to be
\begin{displaymath}
\begin{split}
y_{i} &= (x_{i1},\ldots,x_{ic},x_{j(c+1)},\ldots,x_{jd}) \\
y_{j} &= (x_{j1},\ldots,x_{jc},x_{i(c+1)},\ldots,x_{id}).
\end{split}
\end{displaymath} 
$k$-point crossover works similarly but with $k$ crossover points drawn randomly. Uniform crossover allows each member of $y_{i}$ to be chosen randomly from $x_{i}$ and $x_{j}$, with the corresponding member of $y_{j}$ chosen from the parent not previously used. \par 
Each of these crossover moves is symmetric \citep{liangwong01}, so the only part of the transition probabilities $q_{ij}(\yvec|\xvec)$ that may not cancel in the calculation of the acceptance probability is the probability of choosing $x_{i}$ and $x_{j}$ to cross. We therefore only consider this probability and let $q_{ij}(\yvec|\xvec)=p(I=i|\xvec)p(J=j|\xvec,I=i) + p(I=j|\xvec)p(J=i|\xvec,I=j)$. Then the acceptance probability for crossover is 
\begin{displaymath}
r_{c} = \mathrm{min}\left(\frac{\dens{i}{y_{i}}\dens{j}{y_{j}}q_{ij}(\xvec|\yvec)}{\dens{i}{x_{i}}\dens{j}{x_{j}}q_{ij}(\yvec|\xvec)},1\right)
\end{displaymath}
so that
\begin{displaymath}
r_{c} =
\begin{cases}
\mathrm{min}\left(\frac{q_{ij}(\xvec|\yvec)}{q_{ij}(\yvec|\xvec)},1\right) & \text{if } y_{i} \in \tsubspace{i} \text{ and } y_{j} \in \tsubspace{j}\\
0 & \text{otherwise}.
\end{cases}
\end{displaymath}
We can always choose $p(I=i|\xvec)=p(I=i)=1/\nchains$ and $p(J=j|\xvec,I=i)=p(J=j)=1/(\nchains-1)$ for any $i$,$j$, so that $\yvec$ with $y_{i}\in\tsubspace{i}$ and $y_{j}\in\tsubspace{j}$ is always accepted. However, there may be better choices of these probabilities that ensure the input space is more efficiently explored. In our applications of the algorithm we order the variables in each chromosome so that the most active is first and the least active is last. Using one-point crossover, we then prefer chains closer to the target chain to be crossed with chains further from the target, because this implies that, for crossover point $c$, the first $c$ active variables are roughly in the right area of input space and we explore how these might be combined with choices of the inactive variables that lead to parameter settings passing higher implausibility thresholds. We achieve this by choosing $p(I=i|\xvec)\propto i$ and $p(J=j|\xvec,I=i) \propto (n+1-j)$ for $j\neq i$.
\subsection{Exchange}
The final operation is random exchange, which is the same as the exchange move in a parallel tempering algorithm. We select $i\in\{1,\ldots,\nchains\}$ randomly, then select $j=i\pm1$ with equal probabilities (but with $p(j=\nchains-1 | i=n)=p(j=2|i=1)=1)$. Let $i < j$, relabelling if necessary, then the new proposed state of the Markov chain sees $x_{i}$ and $x_{j}$ swapped so that $\yvec=\{x_{1},\ldots,x_{j},\ldots,x_{i},\ldots,x_{\nchains}\}$. We accept $\yvec$ with probability $r_{e}$ where
\begin{displaymath}
r_{e} = \frac{\dens{i}{x_{j}}\dens{j}{x_{i}}}{\dens{i}{x_{i}}\dens{j}{x_{j}}} =
\begin{cases}
1 & \text{if } x_{i} \in \tsubspace{j} \\
0 & \text{otherwise}.
\end{cases}
\end{displaymath}
\subsection{The implausibility driven evolutionary Monte Carlo algorithm}
The implausibility driven evolutionary Monte Carlo algorithm (IDEMC) is implemented in the following way. Let $\xvec^{(0)} = (x_{0}^{(0)},x_{1}^{(0)},\ldots,x_{\nchains}^{(0)})$ be an initial set of starting points in $\tspace$ with $x_{i}^{(0)}\in\tsubspace{i}$ and $\tsubspace{i}=\{x\in\tspace:|\imp{x_{i}}|\leq\ilevel_{i}\}$ for $i>0$ and $\ilevel_{1},\ldots,\ilevel_{\nchains}$ a well chosen implausibility ladder. The $x_{0}$ chromosome has target distribution $\pi_{0}(x_{0})\sim\mathrm{Uniform}(x\in\tspace)$ and so does not require an implausibility level. When mutating $x_{0}$, we don't require a random walk proposal as we can sample uniformly from $\tspace$ directly. \par 
For $i=1,\ldots,\nchains$ construct a finite partition of $\tspace$, $\tsubspacef{i}{1}\cup\cdots\cup\tsubspacef{i}{r_{i}}$ and set an appropriate variance matrix $\vrw{i}{j}$ for sampling $x_{i}$ from each $\tsubspacef{i}{j}$. For each $\tsubspacef{i}{j}$, select an appropriate variance matrix for sampling from the whole subspace, $\vrw{i}{\whole}$, and let $p_{m}\in(0,1]$ be the mutation rate. Then one iteration of the algorithm proceeds as follows:
\begin{enumerate}
\item Choose mutation or Real crossover with probabilities $p_{m}, (1-p_{m})$ respectively. For mutation, all chromosomes are updated $M$ times each, for some real $M$. For real crossover, select a type of crossover move (one-point, $k$-point or uniform) and select and cross pairs $(\nchains+1)/2$ times.
\item Apply random exchange $\nchains+1$ times as described above.
\end{enumerate}
\par After applying the algorithm for a large number of iterations, the required uniform sample from NROY space is $\{x_{n}^{(t)}, t=1,2,\ldots\}$. IDEMC requires a number of things to be set up before starting; including, most importantly, an implausibility ladder. Choice of the number and setting of the implausibility levels controls the mixing and efficiency of the algorithm in the same way that choosing a temperature ladder effects standard parallel MCMC efficiency. Increasing the number of levels amounts to making the chromosomes sample spaces that are closer together, and so improves mixing. However, the computational burden is increased when more chromosomes are added, so there is a trade off. \par

Also important, but less so than the implausibility ladder, is the way each subspace $\tsubspace{i}$ is sampled during the mutation step through choice of partition $\tsubspacef{i}{1}\cup\cdots\cup\tsubspacef{i}{r_{i}}$, and the variance matrix used to generate proposals from samples within each partition. The more the proposal closely overlays each targeted subspace, the lower the autocorrelation between samples in NROY space will be. This is because it will be easier for new mutations to be accepted. However, an efficient random walk proposal at the mutation step is not as crucial to the performance of the algorithm as the choice of the implausibility ladder. A poor choice of partition will lead to a heavier reliance on the exchange steps to sample NROY space and thus more thinning will be required to obtain a uniform sample even if the implausibility ladder is well chosen. \par 

A final important issue with the initialisation of this algorithm is the requirement that we have start values $\xvec^{(0)} = (x_{0}^{(0)},x_{1}^{(0)},\ldots,x_{\nchains}^{(0)})$ with $x_{j}^{(0)} \in \tsubspace{j}$ for $j=1,\ldots,\nchains$. If NROY space is very small, it may be very difficult to find any points at all satisfying the implausibility criterion for the lower implausibility levels. This can be a real problem in a history matching context as, theoretically NROY space can be empty (we discuss this specific issue in section \ref{discussion}), or it could be so small that it might require trillions of emulator evaluations just to land one point in it.  We present an algorithm in section \ref{ladder.choice} for initialising IDEMC that selects an appropriate implausibility ladder, fits the required parameters for the random walk proposals and finds suitable starting values (assuming NROY space is not empty) for IDEMC.

\section{Constructing an implausibility ladder}\label{ladder.choice}
Suppose we can sample uniformly from $\tsubspace{i}$. Let $\Number{\samples{i}{\nsamples} \in \tsubspace{k}}$ be the number of samples in $\samples{i}{s}=\{x_{i}^{(1)},\ldots,x_{i}^{(\nsamples)}\}$ that are in some subspace, $\tsubspace{k}$, of $\tsubspace{i}$ defined as in (\ref{defn.subspaces}). Then, for $\ilevel_{j} < \ilevel_{i}$ and for a uniform sample from $\tsubspace{i}$, $\samples{i}{\nsamples}$, $\Number{\samples{i}{\nsamples}\in\tsubspace{j}}/\Number{\samples{i}{\nsamples}\in\tsubspace{j}}$ is an estimate of quantity $\pvol$, the ratio of volumes of the two spaces $\tsubspace{i}$ and $\tsubspace{j}$. Note that $\pvol$ is also the probability of making an exchange during step 2 of IDEMC if $j=i+1$ and chromosomes $i$ and $j$ are selected for exchange. Labelling the estimate of $\pvol$, $\hpvol$, the variance of this estimate is $\hpvol(1-\hpvol)/\sqrt{\nsamples}$. \par

We can use these ideas to iteratively set values of $\ilevel_{i}$ during simulation so that $\hpvol$ is within an appropriate tolerance of some pre-chosen desirable value of $p$ between each level. We begin by taking $\nsamples$ uniform samples from $\tspace$, $\samples{0}{\nsamples}$, and set $\ilevel_{1}$ such that $\Number{\samples{0}{\nsamples}\in\tsubspace{1}}/\nsamples \approx \pvol$ (within a tolerance of $\hpvol(1-\hpvol)/\sqrt{\nsamples}$).\par

Next we use $\{\samples{0}{\nsamples}\in\tsubspace{1}\}$ to choose an appropriate partition for $\tspace$, $\tsubspacef{1}{1}\cup\cdots\cup\tsubspacef{1}{r_{1}}$, and set appropriate variance matrices for the random walk proposal $q_{1}(y_{1}|x_{1})$ using a clustering algorithm. First, we perform the clustering algorithm so that, for each member of $\samples{0}{\nsamples}\in\tsubspace{1}$, a cluster is assigned. We then compute $\vrw{1}{j}$, for $j=1,\ldots,r_{1}$, the chosen number of clusters, by computing the empirical variance of the samples within each cluster. The mean, $\mu_{j}$ within each cluster is also retained. We define the partition $\tspace= \tsubspacef{1}{1}\cup\cdots\cup\tsubspacef{1}{r_{1}}$ so that any $x\in\tsubspace{1}$ is in $\tsubspacef{1}{j}$ if
\begin{displaymath}
j = \mathrm{argmin}_{j}\left\{(x-\mu_{j})^{T}\vrw{1}{j}^{-1}(x-\mu_{j})\right\}.
\end{displaymath}
We then set $x_{0}^{(0)}=x_{0}^{(\nsamples)}$ and $x_{1}^{(0)}=x_{0}^{(\mathrm{argmax}_{k}\{x_{0}^{k}\in\tsubspace{1}\})}$. \par
In our implementation of the algorithm we have used k-means clustering with BIC to select the number of clusters (with a pre-chosen maximum number of clusters) at any iteration in order to perform the clustering. However, any clustering algorithm may be used and k-means was chosen here for it's convenience. An example of using a clustering algorithm to tune proposals in MCMC samples was given in \cite{giordanikohn10}, though they use k-harmonic means clustering. \par

Having completed these initial steps we now apply the following algorithm to set the remaining implausibility levels and random walk proposal parameters. Starting at $L=1$
\begin{enumerate}
\item Run IDEMC for $\nsamples$ iterations on the population $\xvec=(x_{0}^{(0)},\ldots,x_{L}^{(0)})$ using starting parameters generated in the previous steps.
\item Let $\samples{L}{\nsamples}=x_{L}^{(1)},\ldots,x_{L}^{(\nsamples)}$ and set $\ilevel_{L+1}$ such that $\Number{\samples{L}{\nsamples}\in\tsubspace{L+1}}/\nsamples \approx \pvol$. If the chosen $\ilevel_{L+1}<a$, set $\ilevel_{L+1}=a$.
\item Use all samples collected in $\tsubspace{1},\ldots,\tsubspace{L+1}$ to define $q_{i}(y_{i}|x_{i})$ for $i=1,\ldots,L+1$ using the clustering method described above.
\item Set $x_{0}^{(0)}=x_{0}^{(\nsamples)},\ldots,x_{L}^{(0)}=x_{L}^{(\nsamples)}$, $x_{L+1}^{(0)}=x_{L}^{(\mathrm{argmax}_{k}\{x_{L}^{(k)}\in\tsubspace{L+1}\})}$. If $\ilevel_{L+1}>a$ return to step 1.
\item Run IDEMC for $s_{n}$ iterations on the population $x_{0}^{(0)},\ldots,x_{\nchains}^{(0)}$.
\item Use all samples collected in $\tsubspace{1},\ldots,\tsubspace{\nchains}$ to define $q_{i}(y_{i}|x_{i})$ for $i=1,\ldots,\nchains$ using the clustering methods stated above.
\end{enumerate}
In practice the number of samples used to perform clustering at steps 3 and 6 can become very large for chromosomes higher up the ladder, so that clustering becomes computationally burdensome. Our solution is to thin the samples so that there is a fixed maximum number of samples used at each stage. This has the benefit of ensuring that the samples used in clustering become more uniform as the burn in process continues. \par The ladder selection algorithm guarantees our IDEMC will have good mixing. The number of chromosomes selected also gives us an estimate of the relative volume of NROY space in relation to $\tspace$. Because each $\tsubspace{i}$ is approximately $\pvol$ times the volume of $\tsubspace{i-1}$, then NROY space is approximately $\pvol^{\nchains-1}$ times the volume of $\tspace$. When we apply IDEMC we first apply the ladder selection algorithm and refer to this as the ``burn in phase''. \par If we have not yet found a point in NROY space before beginning our search technique, there are two options. One is that NROY space has an extremely small relative volume, in which case the ladder selection algorithm will eventually find it; or NROY space may be empty. We address this case as part of the discussion in section \ref{discussion}. We now provide some examples and will define the algorithm for a more general implausibility following multiple waves of history matching in section \ref{multi.wave}.

\section{Illustrative Examples}\label{examples}
We now demonstrate the use of the IDEMC algorithm on several examples, before introducing the main application in section~\ref{application}. These toy examples are chosen to highlight certain features and benefits of such an algorithm, as well as to help clarify the above technical description. We begin with a simple 2-dimensional (2D) example, in order to aid visualisation of the way the chromosomes aid sampling from NROY space.

\subsection{A simple 2D illustration}\label{ssec_2d}

\begin{figure}[ht]
\begin{center}
\begin{tabular}{ll}
\hspace{-0.6cm} \includegraphics[scale=0.56,angle=0]{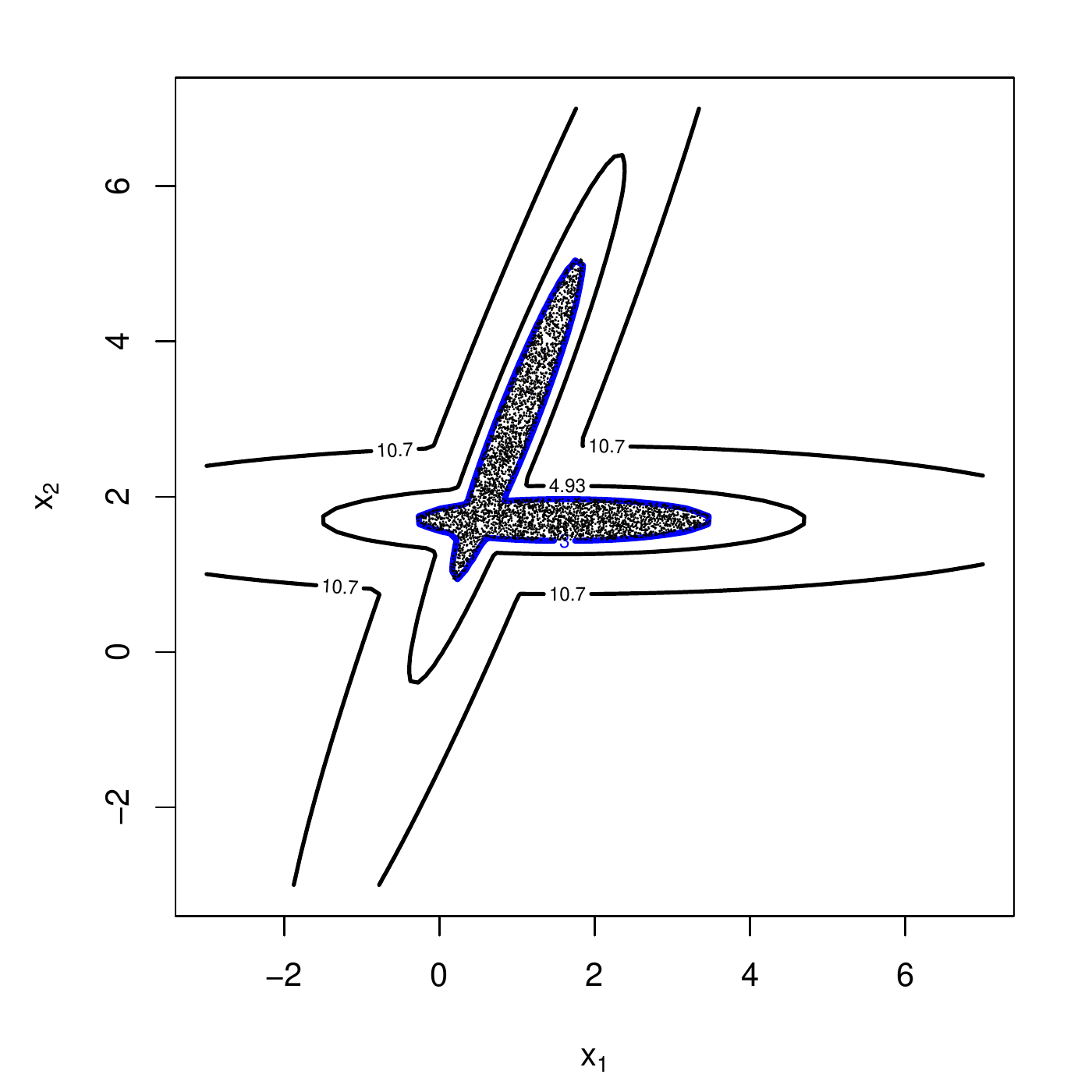}  &
\hspace{-1.0cm} \includegraphics[scale=0.56,angle=0]{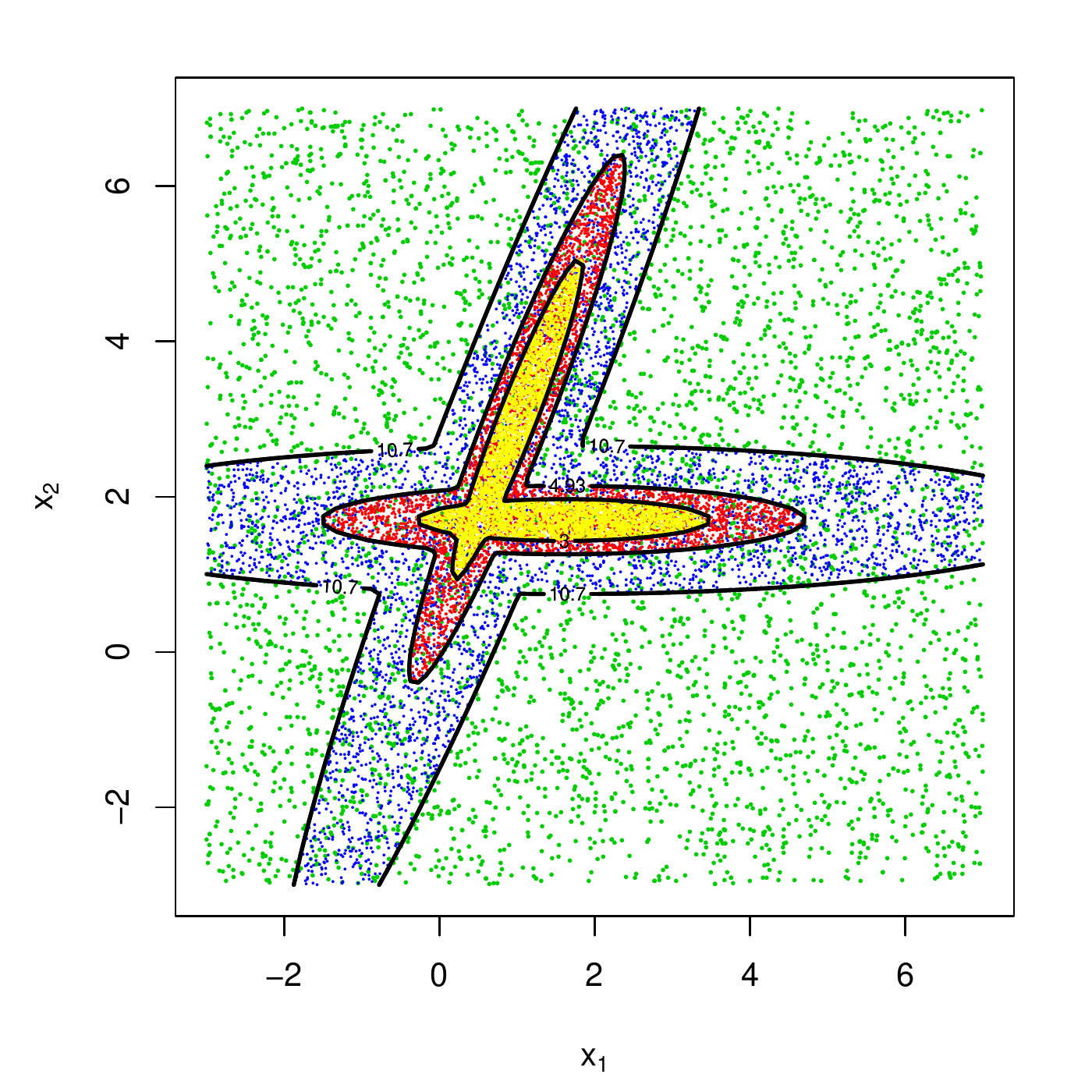}  
\end{tabular}
\end{center}
\caption{Left Panel: the simple 2-dimensional example described in section~\ref{ssec_2d}. During burn-in the IDEMC algorithm selects the implausibility levels, shown here as the black contours. The blue contour defines the boundary of the target region of NROY space (here having $\imp{x} <3$), and the black dots inside this boundary are the uniform draws from the output of the IDEMC. Right panel: the locations of the 5000 steps of each of the four chromosomes (green, blue, red and yellow) with each chromosome residing entirely within its chosen implausibility level. Note the apparent uniformity of all chromosomes.
}\label{fig_2d}
\end{figure}

We begin by exploring the 2-dimensional space $[-3,7]^{2}$ and 
search for the NROY region corresponding to an implausibility measure of the form:
\ba
\imp{x} &=& \min\{A_1(x),A_2(x)\}  \\ 
A_i(x) &=& \sqrt{(x-m_i)^{T}\sigma_{[i]}^{-1}(x-m_i)}, \quad i=1,2
\ea
where  $m_1=(1.6,1.7)^T$, $m_2 = (1,3)^{T}$ and $\sigma_{[i]}$ are  variance matrices with 
\be
\sigma_{[1]} = 
			\left( \begin{array}{cc}
				0.4 & 0 \\
				0 & 0.008
				\end{array} \right)     \quad \quad \rm{and}   \quad \quad
\sigma_{[2]} =
			 \left( \begin{array}{cc}
				0.08 & 0.186 \\
				0.186 & 0.48
				\end{array} \right).
\ee
That is, the implausibility is the minimum of (the square root of) two quadratic forms. We use an implausibility cutoff of 3 which defines the NROY space to be $\tspacef{NROY} =  \{x\in\tspace : |\imp{x}|\leq 3\}$. Note that $\imp{x}$ has a typical characteristic for an implausibility built from a simple computer model: it is the combination of multiple (two in this case) constraints which could have been derived from multiple univariate computer model outputs. 
Figure~\ref{fig_2d}, left panel shows the input space $[-3,7]^{2}$ and the boundary of NROY space is shown as the blue contour.

The IDEMC algorithm is run with $p=0.3$, $M=10$, $p_{M}=0.9$, $s=500$ and $s_{n}=500$ in order to obtain $5000$ samples from the target space. During burn-in the algorithm selected the implausibility levels at 10.7 and 4.93 shown as the black contours in figure~\ref{fig_2d} (left panel), and hence used 
4 chromosomes in total (as there is also the fixed implausibility level of 3 that defines NROY space).
The black dots inside the NROY space in figure~\ref{fig_2d} (left panel) are the 5000 steps from the output of the bottom chromosome of the IDEMC, showing the desired uniformity. 

Figure~\ref{fig_2d} right panel, shows the locations of the 5000 steps of each of the four chromosomes (top to bottom chromosomes are coloured green, blue, red, and yellow respectively), each of which lie entirely within the appropriate implausibility levels. 
Note the apparent uniformity of all chromosomes, a feature that is backed up by the mixing plots given in figure~\ref{fig_2d_mixing} in appendix A, which show excellent mixing at all levels. The colours are consistent with figure~\ref{fig_2d}, right panel. 

Good mixing is of course unsurprising for such a low dimensional example. The target NROY space $\tsubspace{\mathrm{NROY}}$ is not particularly small: approximately 0.032 of the original space, and so is quite easy to hit. We now turn to a significantly more challenging example with multiple modes, and far smaller target.

\subsection{Uniform sampling for a complex 3D shape}\label{sec:3d}

In many applications the implausibility measure $\imp{x}$ may be of a complex form, possibly possessing several modes.
This could lead to the NROY space being composed of disconnected regions, and/or having non-trivial topology. 
We now apply the IDEMC algorithm to a 3-dimensional example that has 4 disconnected modes at the lowest chain, while simultaneously having a torus topology for intermediate chains. The volume of the NROY space in this example is $< 6\times10^{-8}$, much smaller and therefore much more challenging than in the previous 2-dimensional case.

The implausibility function is defined over the 3-dimensional space $[-20,40]^3$ as follows.
For a 3-dimensional input vector $x=(x_1,x_2,x_3)^T$, we define the 2-vector $u$ and the variance matrix $\sigma$ as:
\ba
u = \left( \begin{array}{c}
	(x_1-2)^2 - 3 \\
	(x_2-2)^2 - 3 
	\end{array} \right)  , \quad \quad  \rm{and}  \quad \quad
	\sigma = 
			\frac{1}{2^{12}} \left( \begin{array}{cc}
				1 & -0.97 \\
				-0.97 & 1
				\end{array} \right)  
\ea
and the implausibility function is then given by
\be\label{eq_3d_imp}
\imp{x} = \frac{1}{10} \left( \sqrt{u^T \sigma^{-1} u } + \frac{x_3^2}{0.04^2} \right)
\ee
$\imp{x}$ therefore has 4 modes centred around $x_1= 2\pm \sqrt{3} $, $x_2 = 2\pm \sqrt{3}$ and $x_3=0$. 

\begin{figure}[ht]
\begin{center}
\begin{tabular}{ll}
\hspace{-0.4cm} \includegraphics[scale=0.23,angle=0]{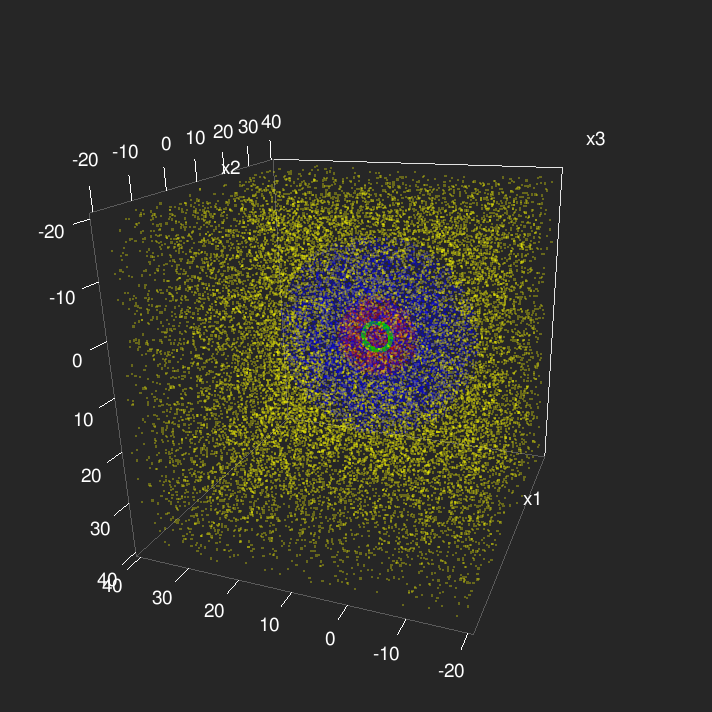}  &
\hspace{-0.4cm} \includegraphics[scale=0.23,angle=0]{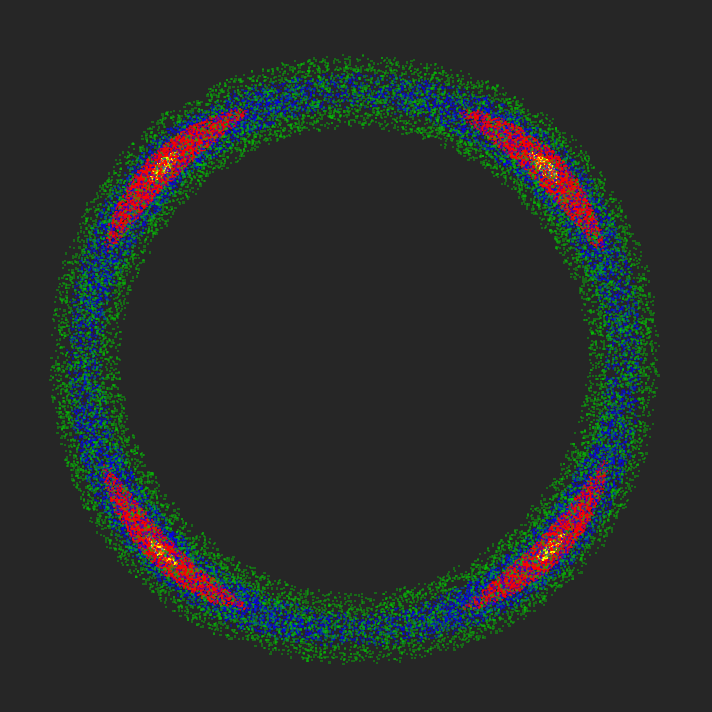}  \\
\hspace{-0.4cm} \includegraphics[scale=0.23,angle=0]{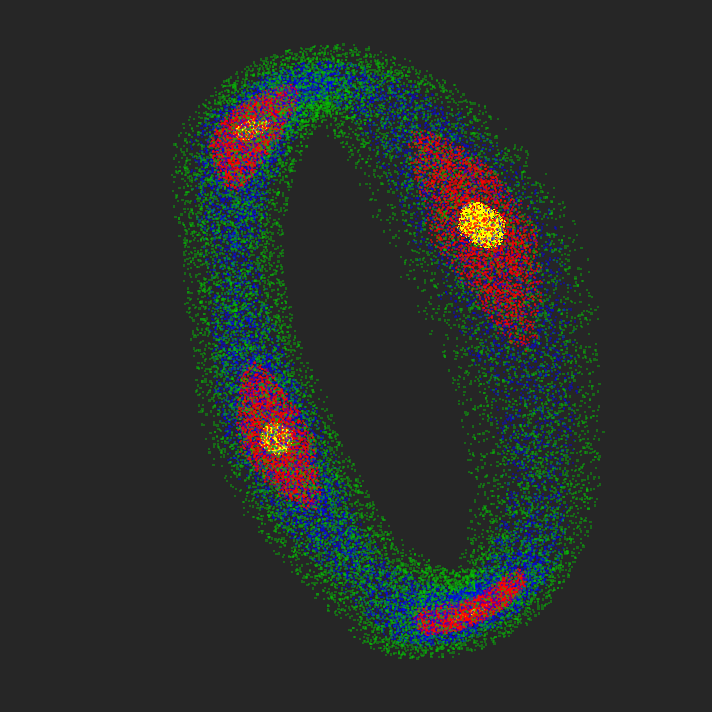} &
\hspace{-0.4cm} \includegraphics[scale=0.23,angle=0]{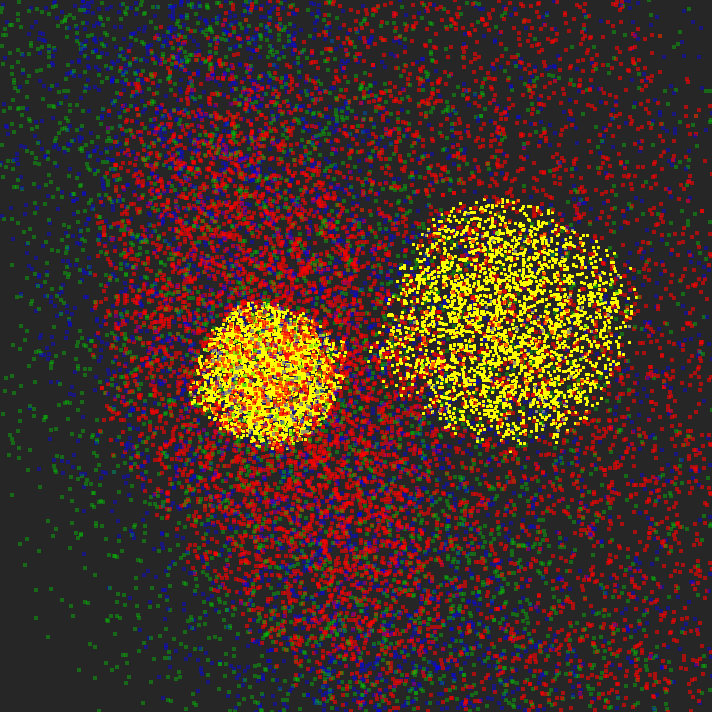}
\end{tabular}
\end{center}
\vspace{-0.5cm}
\caption{The 3-dimensional example described in section~\ref{sec:3d}. Top left: the full input space with samples from chromosomes 
1, 4, 7 and 12 coloured in yellow, blue, red and green respectively. Note the torus topology of chromosome 12. Top right and bottom left/right: zoomed in and rotated views of the green torus corresponding to chromosome 12, along with chromosomes 13, 15 and 20 in blue, red and yellow respectively, the later representing the target NROY space. 
}\label{fig_3d_images}
\end{figure}

The IDEMC algorithm was run with $p=0.4$, $M=15$, $s = 1000$, $s_{n} = 5000$ and $p_{M} = 0.9$ in order to obtain 20000 samples. 20 chromosomes in total were chosen by the algorithm. Figure~\ref{fig_3d_images} shows several images of the results of the IDEMC algorithm for the implausibility given in equation~\ref{eq_3d_imp}.
The top left panel shows the full $[-20,40]^3$ input space with the step locations of chromosomes 1, 4, 7 and 12 coloured in yellow, blue, red and green respectively. 
Note that chromosome 12 takes the form of a green torus: such non-trivial topologies should not pose too serious a problem for the IDEMC algorithm.
Zoomed in and rotated views of the green torus, corresponding to chromosome 12, are shown in the top-right and bottom-left panels of
figure~\ref{fig_3d_images}. Also shown are some of the lower chromosomes: chromosomes 13, 15 and 20 in blue, red and yellow respectively. 
The yellow points therefore lie within the NROY space $\tspacef{NROY}$, again defined using an implausibility cutoff of 3, and the narrowness of the yellow target discs can be clearly seen. 
  
\begin{figure}
\begin{center}
\hspace{-0.6cm} \includegraphics[scale=0.54,angle=0]{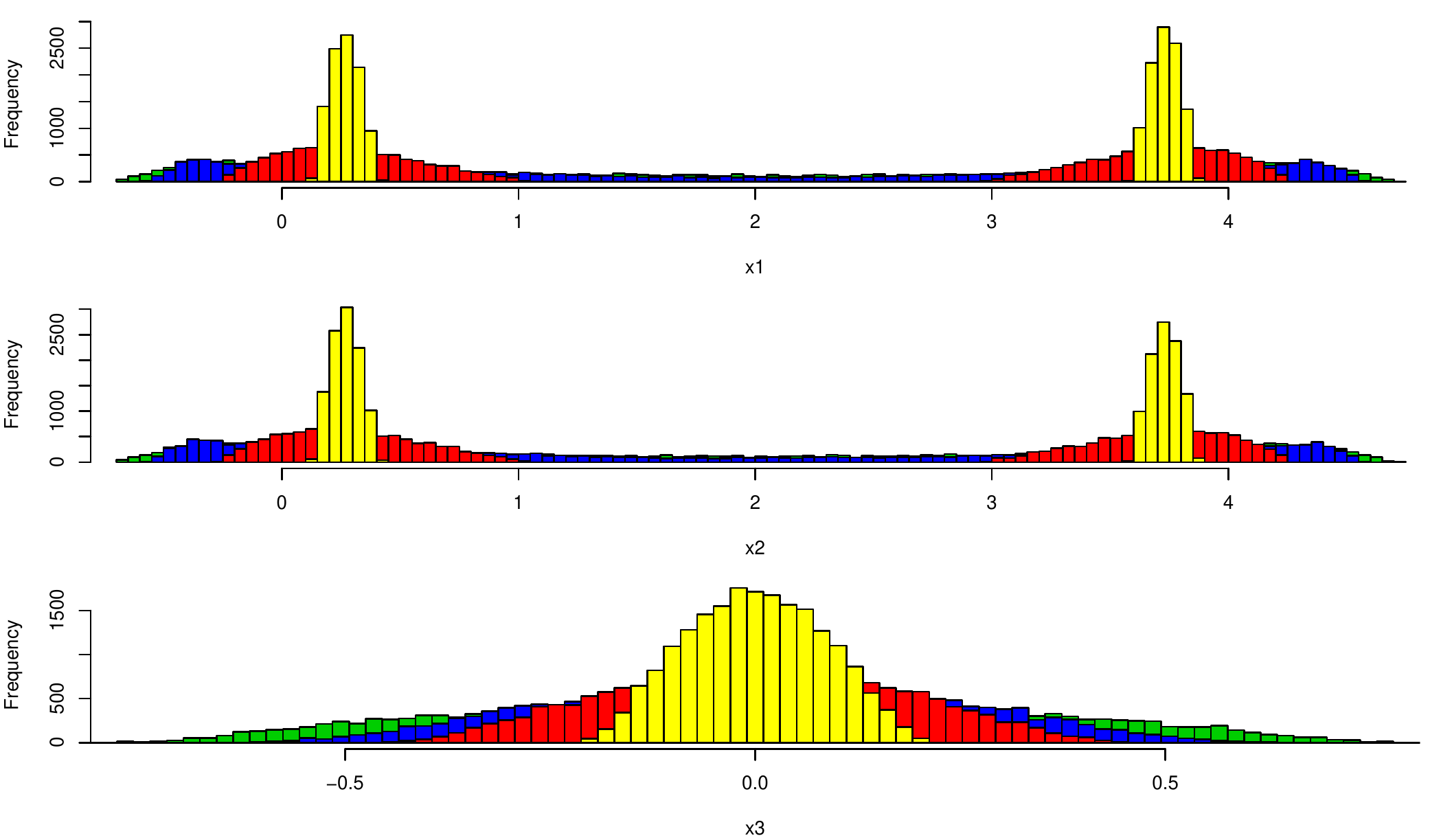}  
\end{center}
\vspace{-0.6cm}
\caption{1-dimensional marginals of the IDEMC algorithm output in the $x_1$, $x_2$ and $x_3$ dimensions (top to bottom panels), for chromosomes 12, 13, 15 and the final chromosome 20, coloured green, blue, red and yellow respectively. The high degree of symmetry in the modes of the $x_1$ and $x_2$ plots (both between plots and within) further demonstrates successful mixing, as does the reflection symmetry about $x_3=0$ in the third plot, as 
bad mixing would most likely lead to noticeable asymmetry. For the full mixing plots see figure~\ref{fig_3d_mixing} in appendix A.
}\label{fig_3d_marginal}
\end{figure}

The bottom-right panel provides a further increase in magnification and shows close up views of two of the 4 modes (viewed by looking through the ring from one mode to its opposite mode). Note the apparent uniformity of the yellow points which is confirmed by the mixing plots, given in figure~\ref{fig_3d_mixing} in appendix A, which again show excellent mixing. One dimensional marginal projections of the draws from chromosomes 12, 13, 15 and 20 are shown in figure~\ref{fig_3d_marginal}, with the same colouring as figure~\ref{fig_3d_images}.
The high degree of symmetry amongst the four modes in the $x_1$ and $x_2$ marginal plots, and the reflection symmetry in the $x_3$ plot again give strong evidence for excellent mixing, as unacceptable mixing in this example would most likely lead to noticeable asymmetries. 

We now go on to examine a higher dimensional example, where the target NROY space is an extremely small proportion of the original input space.


\subsection{A tiny subspace in 10 dimensions}\label{sec:10d}
We now test the capabilities of IDEMC on a 10D example where NROY space is a tiny disconnected subspace of the original 10D box it is defined for. We define an implausibility function within the hypercube defined on $[-3,7]^{10}$ with implausibility $\imp{x} = \min\{A_1(x),A_2(x)\}$ with
\begin{displaymath}
A_i(x) = \sqrt{(x-m_i)^{T}\sigma_{[i]}^{-1}(x-m_{i})}
\end{displaymath}
and $\sigma_{[i]_{jk}} = \gamma^{2}\sqrt{v_{ij}}\sqrt{v_{ik}}C_{jk}$, where $C_{jk}=0.85+0.15\indicator_{j=k}$  
\begin{displaymath}
\begin{split}
m_{1} = (1\ 1\  1\ 1\ 1\ 1\ 1\ 1\ 1\ 1)^{T} &\qquad v_{1} = (0.1\ 0.0125\ 0.025\ 0.04\ 0.01\ 0.1\ 0.0125\ 0.025\ 0.04\ 0.01)^{T} \\
m_{2} = (4\ 3\ 3\ 4\ 3\ 4\ 4\ 4\ 2\ 2)^{T} &\qquad v_{2} = (0.025\ 0.1\ 0.01\ 0.01\ 0.05\ 0.025\ 0.1\ 0.01\ 0.01\ 0.05)^{T},
\end{split}
\end{displaymath}
and $\gamma=0.5838968$ is a scaling factor chosen so that the volume of the target space (where $\imp{x}\leq3$) is $1\times10^{-18}$ (or a billionth of a billionth) of the original parameter space. The target space is the volume inside two extremely small ellipsoids, though some of the chromosomes in an IDEMC for sampling this space will be sampling from the interior of the boundary of the intersection of two ellipsoids when the implausibility cut off is large enough.
\par
We constructed the implausibility ladder and required proposal transition matrices using the algorithm described in section \ref{ladder.choice} for a ratio of volumes between each sampled subspace of $0.3$ with $s=2000$ and $s_{n}=5000$. This led to a population with $36$ chromosomes at levels starting at $226.3$ for the second chromosome (the first samples uniformly over the full space), down to level $3$ for the target chromosome. We then run IDEMC for 100,000 iterations, thinning every $10$th to generate $10,000$ uniform samples from the defined region. 

\begin{figure}[ht]
\centering
\includegraphics[height=0.43\textheight,width=0.65\textwidth]{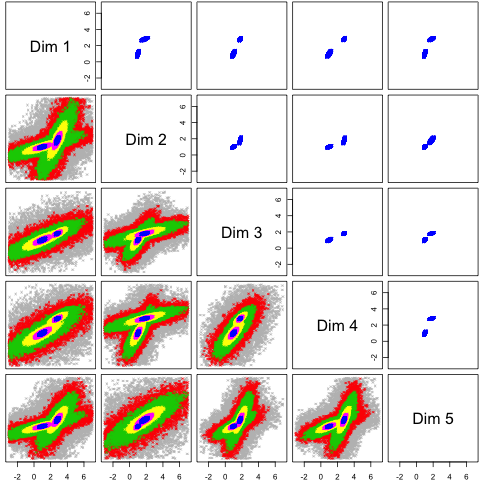}
\caption{2D projections for the uniform sample from 10D target space (upper triangle). The lower triangle shows 2D projections of samples from other chromosomes sampling regions with higher implausibility. Specifically, we chose levels 131 (grey points), 65.5 (red points), 35.3(green points), 15.1 (yellow), 6.5 (magenta) and the target (blue).}\label{pairs10d}
\end{figure}
\begin{figure}[htb]
\centering
\includegraphics[height=0.4\textheight,width=\textwidth]{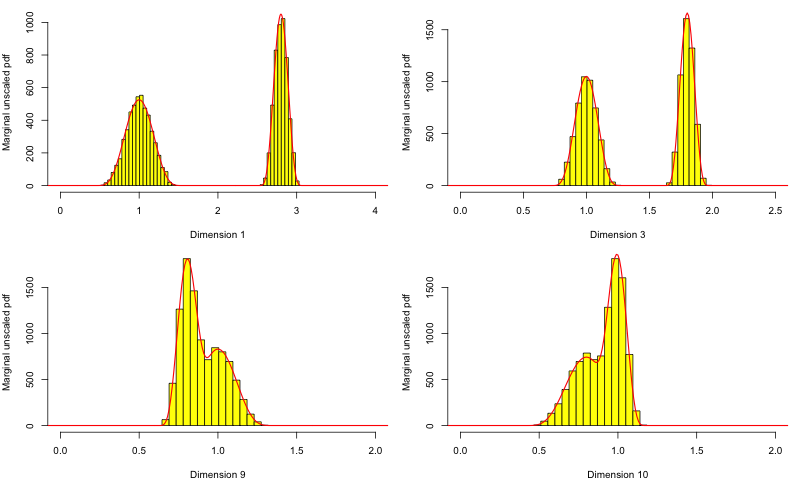}
\caption{Histograms of the uniform samples for 4 variables in the 10-dimensional example with theoretical marginal distributions (red lines) overlaid.}\label{marginals10d}
\end{figure}
Figure \ref{trace10d} (appendix A) shows trace plots for 10 dimensions of the target chromosome and highlights excellent mixing. Figure \ref{pairs10d} shows 2-dimensional projections of the uniform sample of the target space for the first 5 dimensions. The upper triangle shows only the target samples whereas the lower triangle plots samples from other chromosomes sampling regions with higher implausibility to give an idea of the difference between the space defined by each implausibility level. Specifically, we chose levels 131 (grey points), 65.5 (red points), 35.3(green points), 15.1 (yellow), 6.5 (magenta) and the target (blue). The samples from each level are plotted on top of one anther (highest level first). \par 
We demonstrate the quality of the sample by plotting histograms for 4 of the 10 marginal distributions, overlaying their theoretical marginals as red lines. We can see that the samples match closely to the theoretical underlying marginal distributions. This example illustrates the ability of IDEMC to get uniform samples from extremely small, disconnected, subspaces of relatively high dimensional parameter spaces. In order to obtain these $10,000$ uniform samples, we required $1,000,000$ evaluations of the implausibility function after the burn in phase, which itself required $1000+20,000*35+50,000 = 751,000$ runs. Hence, for the cost of $1,751,000$ evaluations of the implausibility function we are able to obtain 10,000 uniform samples of the target space. Whereas, due to the relative size of the target space, we would expect to need $10^{18}$ evaluations to obtain just one sample by rejection sampling.  We discuss the efficiency of our algorithm further in section \ref{al.efficiency}.
\section{Multi-wave implausibilities and empty NROY space}\label{multi.wave}
\subsection{Multi-wave implausibility}
Suppose we have completed $m$ waves of history matching so that we have defined $m$ implausibility measures $\impf{[k]}{x}$ for $k=1,\ldots,m$. Suppose we have defined $m$ implausibility cutoffs $a_{k}$, one for each wave $k=1,\ldots,m$. Then NROY space is $\tsubspace{m}$ with
\begin{displaymath}
\tsubspace{\mathrm{NROY}}=\{x\in\tspace: |\impf{[k]}{x}|\leq a_{k}, k=1,\ldots,m\}
\end{displaymath}
In this situation IDEMC is the same but the implausibility ladder must be more carefully chosen, and the target subspaces must be defined more carefully. This is because to define the implausibility in a naive way, for example, by simply maximising over all implausibilities over each wave, could result in an extremely badly behaved implausibility surface with many disconnected regions. We hence define the implausibility ladder as an $n\times m$ matrix $B$ with $B_{ni} = a_{i}$, $B_{ij}\geq B_{kj}$ for $k>i$ and $B_{ij}=\infty$ if $B_{ik}>3$ for $j=k+1,\ldots,m$. The subspaces of $\tspace$ defined by these levels are
\begin{displaymath}
\tsubspace{i} = \{x\in\tspace:|\impf{[j]}{x}|<B_{ij}; j=1,\ldots,m\}
\end{displaymath}
and $\tsubspace{n}=\tsubspace{\mathrm{NROY}}$.
The ladder selection algorithm works in the same way but with each row of $B$ fixed at each step of the algorithm.

\subsection{Algorithm efficiency}\label{al.efficiency}
IDEMC samples from many different spaces in order to target NROY space. We can compare its efficiency, in terms of the number of emulator evaluations, with current rejection sampling methods as a function of the relative volume of the target space, $V_{T}$, and of some of the other key parameters used in the algorithm. We can then use this to decide when IDEMC should be preferred to rejection sampling. \par Let $N$ be the number of samples from NROY space required, $N_{e}^{r}$ the number of emulator evaluations needed to obtain $N$ samples via rejection sampling, and $N_{e}^{A}$ the number required using IDEMC, then, clearly
\begin{displaymath}
\E{N_{e}^{r}} = \frac{N}{V_{T}}.
\end{displaymath}
The following calculation assumes that $\hat{n} = 1 + \lceil\log{V_{T}}/\log{p}\rceil$ is an accurate estimator for $n$. \par If IDEMC requires $n$ chromosomes then the burn in phase will require $ns + ns_{n}$ MCMC steps, and the post burn in phase requires $nNT$ steps if the sample is thinned every $T$ steps. Remembering that the first chromosome can be sampled from directly during the algorithm because it refers to the full parameter space (so that mutation steps need only be applied to $n-1$ of the chromosomes), then,
\begin{equation}\label{efficiency}
\begin{split}
\E{N_{e}^{A}} = &\quad s\left(1 + \sum_{k=2}^{\hat{n}-1}\left(p_{M}((k-1)M+1) + (1-p_{M})(k+1)\right)\right) \\ &+ s_{\hat{n}}\left(p_{M}((\hat{n}-1)M + 1) 
+ (1-p_{M})(\hat{n}+1)\right) \\ &+ NT\left(p_{M}(M(\hat{n}-1)+1) + (1-p_{M})(\hat{n}+1)\right)
\end{split}
\end{equation}
In our applications of this algorithm to date, we have found $s=2000$ and $s_{n}=5000$ are reasonable choices in higher dimensional spaces that lead to IDEMC's that mix well. Suppose, additionally, that we fix the ratio of volumes between the spaces sampled by adjacent chromosomes via $p=0.4$ so that we have good mixing between the chromosomes, then we can compare $\E{N_{e}^{A}}$ with $\E{N_{e}^{r}}$ for different values of $V_{T}$ and different values of $M$ and $T$, to discover how small the volume of the target space should be before implementation of IDEMC becomes more efficient than rejection sampling. 
\begin{figure}[htb]
\centering
\includegraphics[height=0.4\textheight]{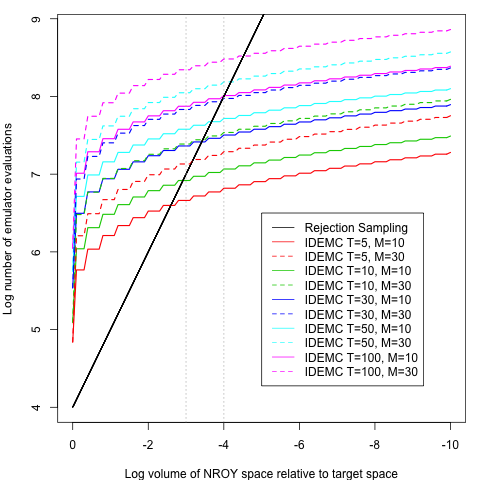}
\caption{$\log{v_{T}}$ plotted against the log of the expected number of emulator evaluations for rejection sampling required to obtain $10,000$ uniform random samples from NROY space (solid black line) and for IDEMC to obtain the same number of uniform draws with T and M as given in the key and with the other parameters fixed as in the text.}\label{efficiency.plot}
\end{figure}
\par Figure \ref{efficiency.plot} plots both $\E{N_{e}^{r}}$ (solid black line) and $\E{N_{e}^{A}}$ for a number of different values of $T$ and $M$ (coloured lines) against $V_{T}$ (both on a $\log_{10}$ scale), assuming that 10,000 uniform random samples are required. The plot shows that if the algorithm mixes well so that not a great deal of thinning is required, IDEMC is more efficient than standard rejection sampling in spaces of around $0.1\%$ of the original parameter space. Even with more thinning, by the time the space is at $10^{-4}$ times the original space or smaller, IDEMC will be more efficient. For the sort of sizes of NROY space we have encountered and that motivated this research, where the space is order $10^{-6}$ times the size of the original space, IDEMC is extremely efficient. In fact, if the space is $10^{-6}$, IDEMC  is $100$-$1000$ times more efficient than rejection sampling depending on how much thinning is required. \par For small NROY spaces, getting any samples at all by rejection sampling is difficult. For spaces of the order $10^{-6}$ times the original space, we have argued that IDEMC is far more efficient, however, it is still feasible to perform rejection sampling. However, for spaces much smaller than this, it will not be feasible. We have encountered such spaces in our own research. If NROY space is $10^{-6}$ times the size of the original space, most of the samples we obtain from it will have implausibility between $2$ and $3$. Though all of these samples cannot be ruled out as possible values of $x^{*}$, it is often of interest to look at values in NROY space with implausibility less than $2$ or even $1$. This is particularly true after multiple waves of history matching, when the emulators are quite accurate so that low values of implausibility are more likely to be due to good matches rather than large emulator variances or inaccuracies. These subsets of NROY space with low implausibilities can be extremely small, even relative to NROY space and may only be feasibly sampled using an algorithm such as IDEMC.

\section{Application}\label{application}

\subsection{Introduction to Galaxy Formation}

The fundamental goal of Cosmology is the understanding of the evolution of the Universe, from the Big Bang to the current day.
A critical part of this is the study of structure formation: cosmologists wish to understand the formation, subsequent growth and evolution of millions of observed galaxies in the presence of Dark Matter. 
A state of the art ``semi-analytic" computer model of galaxy formation has been constructed by the world-leading GALFORM group, based at the Institute of Computational Cosmology, Durham University, UK (see \cite{bower06} and references therein). 
GALFORM simulates the creation and evolution of millions of galaxies and hence can be used to test the state-of-the-art theories of cosmology and of galaxy formation. 
Like many computer models it takes a significant time to run, and has many input parameters to explore. 
The GALFORM model can also produce many outputs that can be compared to real world data, but arguably one of the most important of these is known as the Stellar Mass Function \citep{bower12}.

By determining whether the latest version of the GALFORM model can match current observations such as the stellar mass function, cosmologist learn more about 
proposed physical processes thought to govern structure formation \citep{bower10}. Therefore, it is of most scientific interest to confirm if GALFORM can match these outputs, and to determine for which locations in input parameter space such matches can be found. 
Note that at this stage, a fully Bayesian calibration of the input space is inappropriate, as we are in no way certain that the model can even match the data with sufficient accuracy to warrant such a detailed study. 
Instead, in this scenario, a history match analysis is appropriate, where unlike in a Bayesian calibration we do not assume the explicit existence of a `best input' $x^{*}$. This allows for the possibility of ruling out the whole input space, if the model is unable to match the observed data within the specified tolerances. See \cite{vernonetal10} (and rejoinder) for further discussions of the comparison between these approaches and \cite{StatSci13} for an overview of history matching as applied to galaxy formation simulations.

Four waves of emulation were performed on the GALFORM model (see~\cite{Rodrigues13} for details of the latest version of GALFORM used in this study), and we now go on to 
describe the current state of this history match analysis. We then apply the IDEMC algorithm to generate a uniform design of runs over the current wave 4 NROY space, for use in the next wave of analysis.
\begin{figure}[t]
\begin{center}
\begin{tabular}{ccc}
\hspace{-1.5cm}
\includegraphics[scale=0.34,angle=0,viewport= 27 58 600 425,clip]{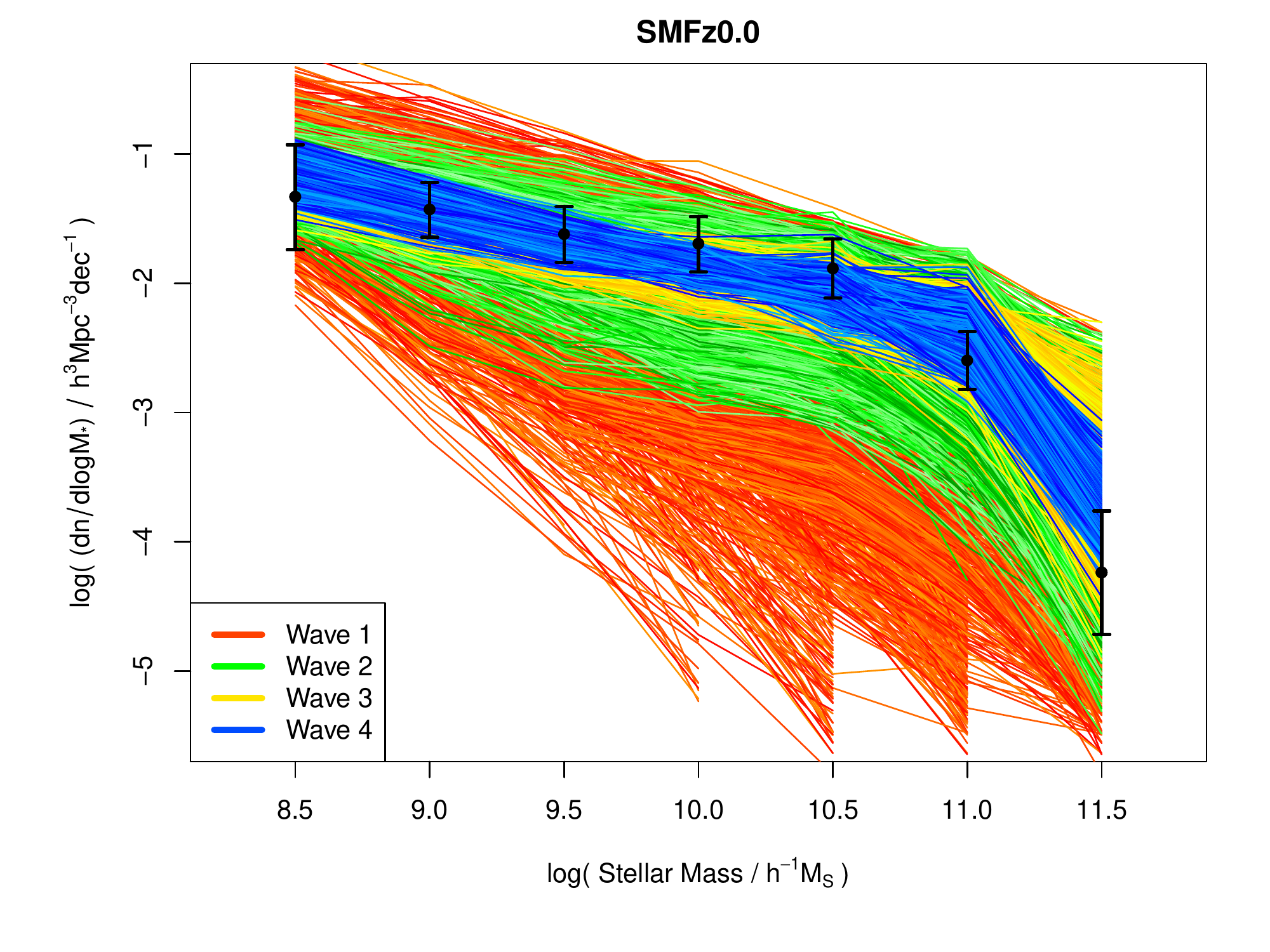} & \hspace{-1.0cm}
\includegraphics[scale=0.34,angle=0,viewport= 60 58 600 425,clip]{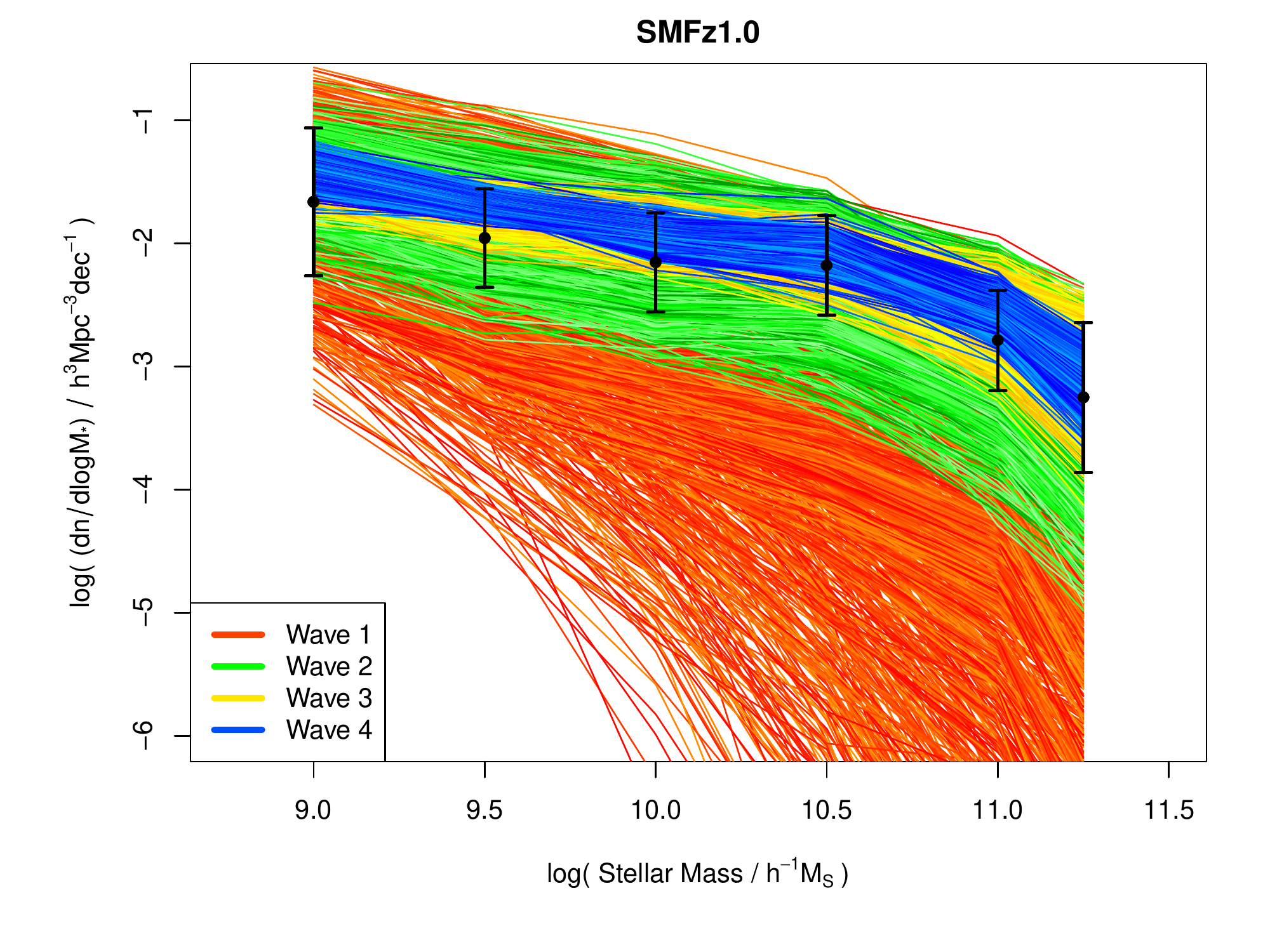} & \hspace{-1.0cm}
\includegraphics[scale=0.34,angle=0,viewport= 60 58 600 425,clip]{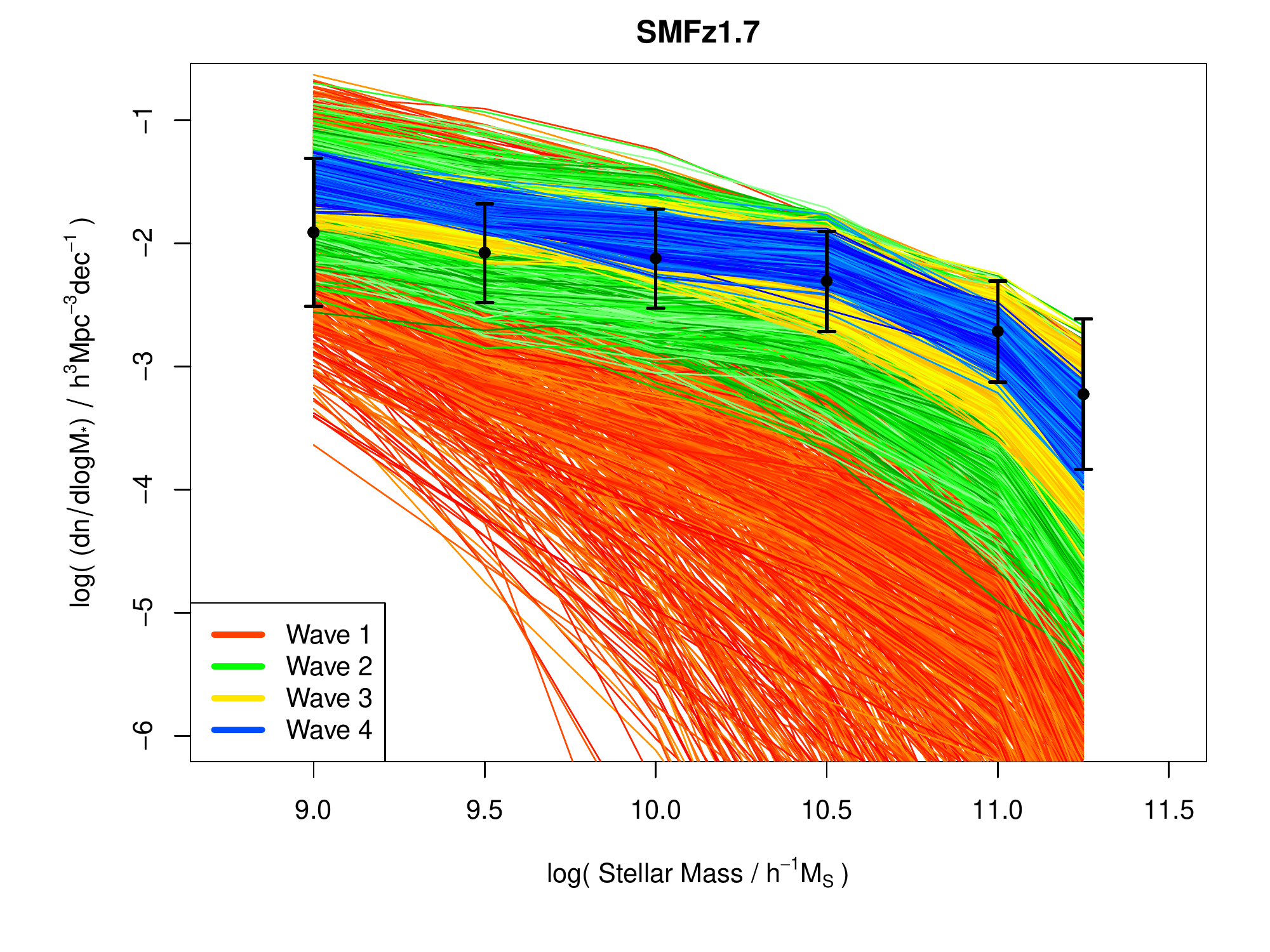} \\
\multicolumn{3}{c}{
\hspace{-1.5cm}
\includegraphics[scale=0.34,angle=0,viewport= 27 28 600 425,clip]{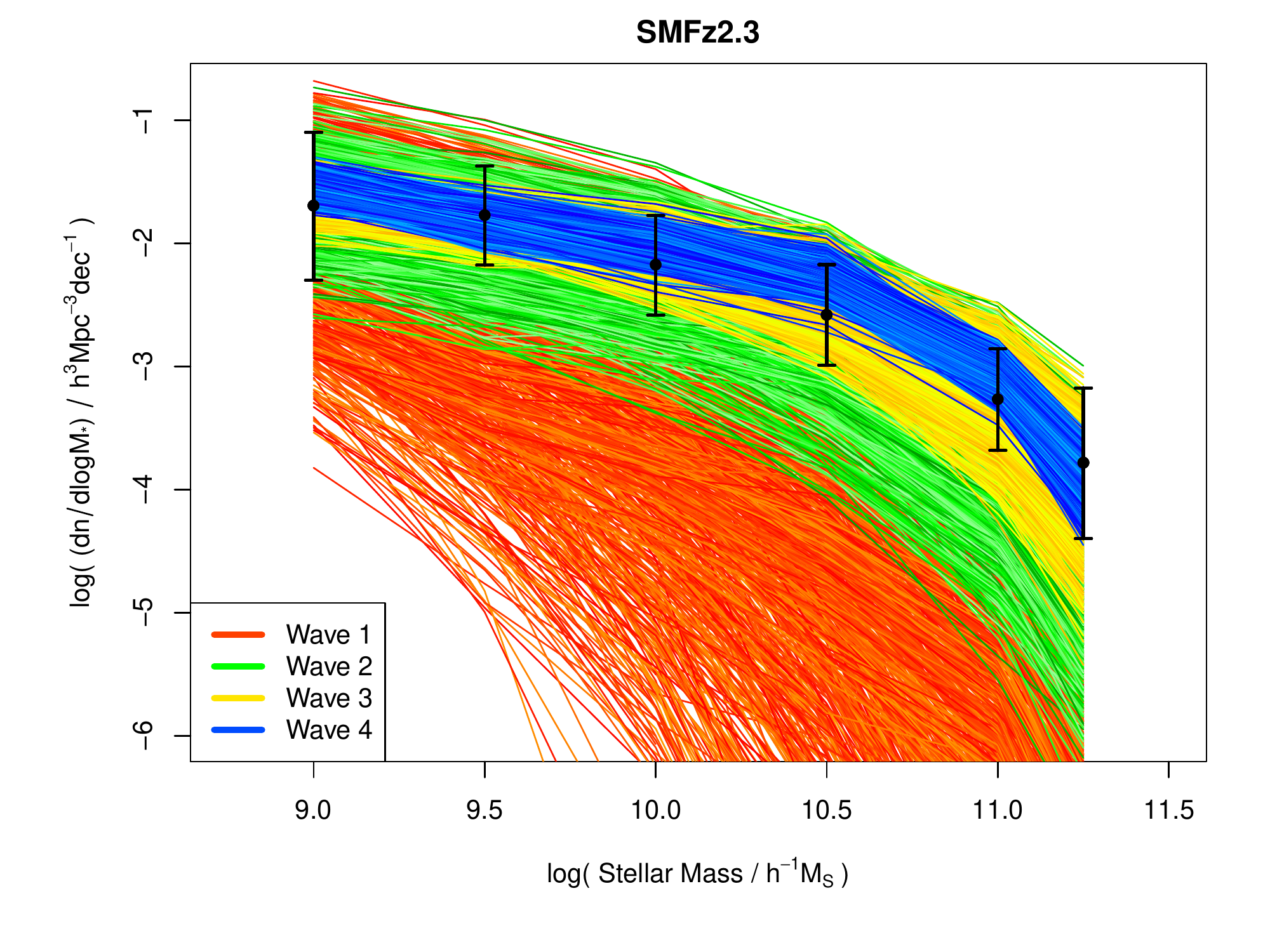}   \hspace{-0.6cm}
\includegraphics[scale=0.34,angle=0,viewport= 60 28 600 425,clip]{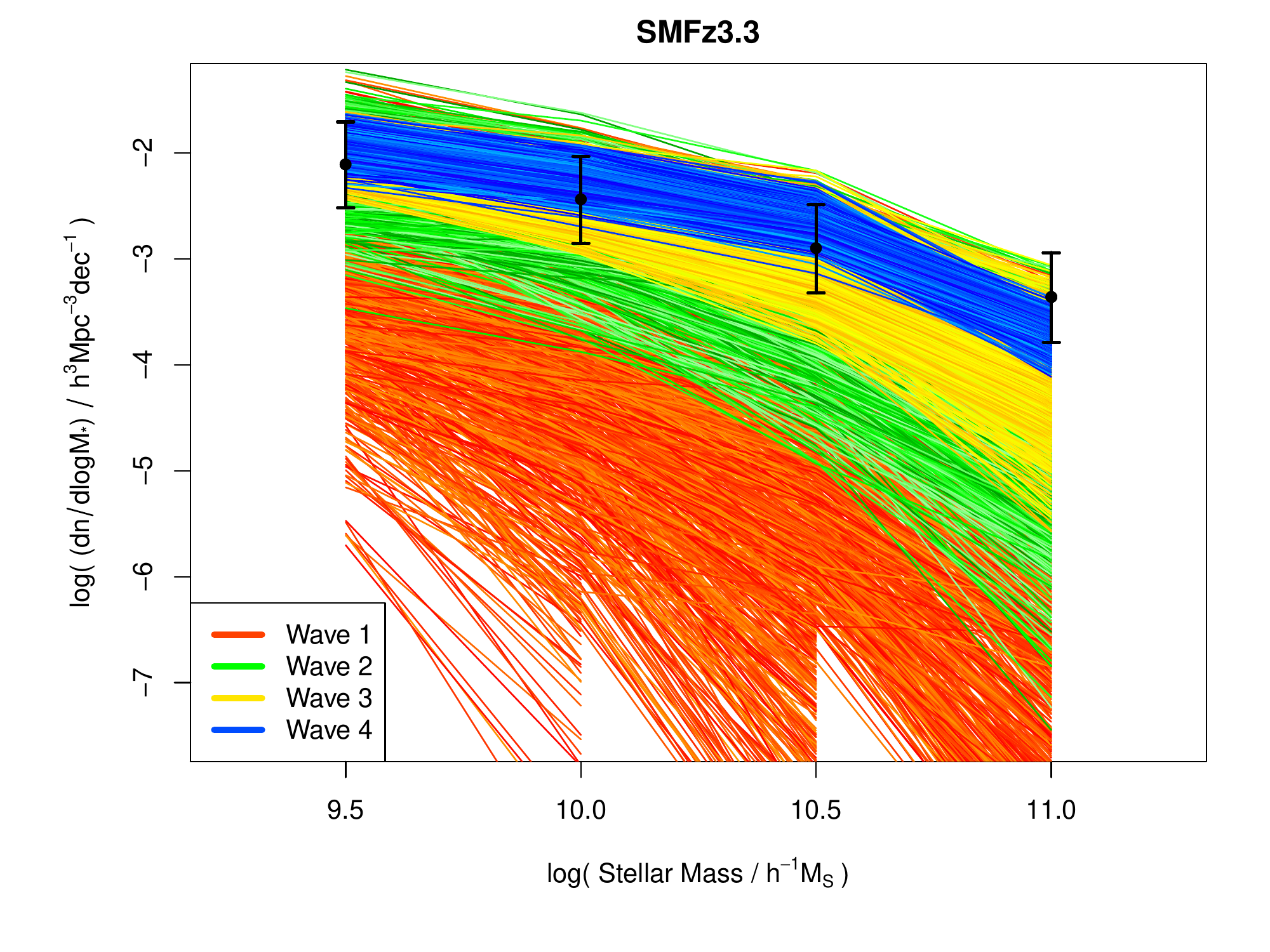} 
}
\end{tabular}
\caption{The main GALFORM outputs considered: the Stellar Mass Functions at redshifts 0, 1, 1.7 2.3 and 3.3 (top left to bottom right panels). Higher redshift corresponds to more distant galaxies and hence, due to the finite speed of light, to observations made further in the past. The black points are the observed data, while the $\pm$ 3 sigma  error bars represent the observation error plus the expert elicited model discrepancy (see equations (\ref{observations}) and (\ref{best.input})). The runs from waves 1 to 4 are shown as the red, green, yellow and blue lines respectively.
}\label{fig_galf_outs}
\end{center}
\efi
\hspace{-1cm}
\begin{table}
\resizebox{\textwidth}{!}{%
\begin{tabular}{|c|c|c|c|c|c|c|c|}
\hline
Input &min&max &Process & Input &min&max &Process \\ 
Parameter & & & Modelled & Parameter & & & Modelled \\
\hline
 $\beta_{\rm 200,disk}$    & 0.5 & 40 & SNe feedback 	&  $\yield$				&0.02&0.05 & Metal Enrichment	\\
$ \beta_{\rm 200,burst}$  & 0.5 & 40 & $\cdot$ 	   	&  $\VCUT$				&20&60  		& Reionisation\\
$\alphahot$			&1&3.7 & $\cdot$  			&  $\ZCUT$				&5&15 		& $\cdot$  \\
$\alphareheat$			&0.15&1.5 & $\cdot$   		&  $\nu_{sf}$				&0.025&1 	&  Star Formation \\
$\fellip$				&0.01&0.5 &  Galaxy Mergers	&  $P_{\rm sf}/k_{\rm B}$					&10000&50000& $\cdot$  \\
$\fburst$				&0.01&0.5	&  $\cdot$		&  $\beta_{\rm sf}$		&0.65&1.1	& $\cdot$  \\
$\stabledisk$			&0.61&1.1 & Disk stability  	&  $f_{\rm dyn}$			&1&100	 	&  $\cdot$  \\
$\alphacool$			& 0.1& 2 	&  AGN feedback	&   $\tau_{\rm star}$			&0.001&0.5	& $\cdot$  \\
$\epsilonSMBHEddington$&0.004&0.1 & $\cdot$  		&  &&& \\
\hline
\end{tabular}}
\caption{The 17 input parameters to the GALFORM model that make up the vector $x$ and their associated ranges (which were converted to -1 to 1 for
the analysis). The input parameters are grouped by their relevant physical process. 
}\label{tab_inputs}
\end{table}



\subsection{Current Status of the History Match Analysis}

The version of GALFORM analysed within this study has 17 input parameters which relate to various physical processes involved in galaxy formation (see~\cite{vernonetal10} for more details). The input parameters and their relevant physical processes are shown in table~\ref{tab_inputs} along with the ranges that define the boundaries of the input space.
The cosmologists are interested in matching 5 stellar mass functions, with each function possessing between 4-7 individual outputs, as shown in figure~\ref{fig_galf_outs}. The 5 stellar mass functions each relate to a specific redshift or time: higher redshift relates to observations that are at larger distances and hence further back in time. The outputs within each mass function represent the number of galaxies observed for given stellar mass~\citep{bower12}. The black points in figure~\ref{fig_galf_outs} represent the observational data~\citep{Mortlock11}, and the error bars represent $\pm$ 3 sigma and contain both the observational errors and the elicited model discrepancy as given in equations~(\ref{observations}) and (\ref{best.input}) respectively. See \cite{imprecise} for discussions of the elicitation of model discrepancies in the context of galaxy formation.

The current state of the history match is that 4 waves of emulation have been completed, following the methodology described in~\cite{vernonetal10} (also see section~\ref{history.matching}). At each wave a total of 5000 runs were performed within the current NROY space (with only 3800 runs for wave 1 due to technical problems). The stellar mass function outputs of the runs generated in waves 1 to 4 are shown in figure~\ref{fig_galf_outs} as the red, green, yellow and blue lines respectively. At each wave, emulators were constructed for a subset of the 29 outputs using equation~(\ref{emulator}), and for each output considered an implausibility measure $\impf{i}{x}$ was built. Various combinations of the $\impf{i}{x}$ were used to reduce the input space: we used the second and third maximum of $\impf{i}{x}$ for early waves (denoted $\impf{2M}{x}$ and $\impf{3M}{x}$ respectively), and included the first maximum $\impf{M}{x}$ in later waves when we were more confident about emulator accuracy.

Table~\ref{tab:waves} shows the implausibility measures used so far, and the corresponding cutoffs and the proportion of input space remaining as NROY after each wave. It also shows the number of outputs emulated, the number of inputs identified as active and the order of the polynomial terms used in the linear model part of the emulator. 
At each wave, various emulator diagnostics were performed, and conservative implausibility cutoffs chosen appropriately, following the methodology described in~\cite{vernonetal10}. The challenge is now to generate a uniform sample from the wave 4 NROY space.

\begin{table}
\begin{center}
\resizebox{\textwidth}{!}{%
\begin{tabular}{|c|c|c|c|c|ccc|c|}
\hline
Wave & Runs & Outputs Emul. & Active Inputs & Poly. Order & $\mathcal{I}_{M}$ & $\mathcal{I}_{2M}$& $\mathcal{I}_{3M}$ &  NROY Space\\
\hline
 1 & 3800 & 10 & 17  & 2nd & - & 3 & 2.4 &   9.62$\times 10^{-2}$   \\
  2 & 5000 & 10 & 10 & 3rd & - & 3 & 2.4  &  9.89$\times 10^{-3}$  \\
 3 & 5000 & 8 & 10  & 3rd & 3.5 & 3 & 2.4  &    3.37$\times 10^{-4}$  \\
 4 & 5000 & 8 & 10  & 3rd & 3.5 & 3 & 2.4  &       1.34$\times 10^{-5}$  \\
\hline  
\end{tabular}}
\end{center}
\caption{Summary of the current state of the history match. Col. 2: the no. of
model runs used to construct the emulators at each wave; Col. 3: no. of outputs emulated, Col. 4: the no. of Active Variables used; Col. 5 the order of polynomial used in the linear model part of the emulator, Col. 6-8: the implausibility thresholds imposed which define the NROY space  at each wave;
Col. 9: the proportion of the original parameter space deemed NROY.} \label{tab:waves}
\end{table}

\subsection{Applying the IDEMC algorithm to GALFORM}\label{ap.to.gal}
We apply IDEMC to GALFORM, constructing the implausibility ladder during the burn in phase by specifying a ratio of volumes between adjacent chromosomes of $0.4$. We allow the probability of crossover to be 0.15 during the burn in phase and allow 10 mutations during a mutation step. We let $s$ from the ladder selection algorithm be 2000 and let $s_{n}$ be $5000$. The constructed implausibility ladder had 14 chromosomes and we found that the volume of NROY space was $1.34\times10^{-5}$ of the volume of the original parameter space.  \par We found that crossovers were very rarely accepted for chromosomes sampling from spaces closer to NROY space. Therefore, though crossover is useful for locating NROY space during burn in, we find it less useful for sampling from the located space. When we run the algorithm during the sampling phase, we therefore lower the probability of crossover to $0.03$. We run the IDEMC algorithm for $150000$ iterations, thinning every $30$ samples in order to generate $5000$ uniform samples from NROY space. The mixing plots for this run are shown in figure \ref{fig_galf_mixing} and show that the algorithm is mixing well. We don't discard any burn in here as all of the runs completed during the burn in phase are discarded. \par Computing $\E{N_{e}^{A}}$ in (\ref{efficiency}) gave an expected number of emulator evaluations for this sample as $21,075,000$. IDEMC is, therefore, roughly 18 times more efficient in terms of number of required emulator evaluations to obtain the sample, than rejection sampling in this application. 

\subsection{Results}

The application of the IDEMC algorithm to the GALFORM history match has been very useful for several reasons. 
Primarily, it has produced 5000 uniform draws from the NROY space defined after 4 waves of history matching. 
These 5000 points are useful both in supplying candidate points for future designs (e.g. for the next wave), and for use in analysing the 
location and geometry of the current NROY space. 
Investigating the shape of the NROY space is often of vital interest to the cosmologists as it aids understanding of the structure 
of the GALFORM model, and is often helpful in identifying current model deficiencies and suggesting improvements for future versions.

Figure~\ref{fig_galf_impdep} shows, in the plots above the diagonal, several 2-dimensional `optical depth' plots of the wave 4 NROY space constructed using the 5000 uniform draws, and corresponding to the 10 most active inputs to the GALFORM model (see~\cite{vernonetal10} for a detailed description of such plots). The optical depth plots are designed to highlight the geometry of the NROY space. They give the depth or volume of the high-dimensional NROY space, conditioned on the 2 inputs given on the x- and y-axis of each plot.\footnote{They hence give the $d-2$ dimensional `depth' of the NROY space were one to look along the direction perpendicular to the 2 conditioning inputs, hence the term optical depth.}
In this case, these plots are therefore proportional to the 2-dimensional marginal distributions of the uniform distribution over NROY space, scaled by its total volume.
The corresponding 1-dimensional optical depth plots are also shown down the diagonal of figure~\ref{fig_galf_impdep}.

\bfi
\begin{center}
\includegraphics[height=0.7\textwidth,angle=90]{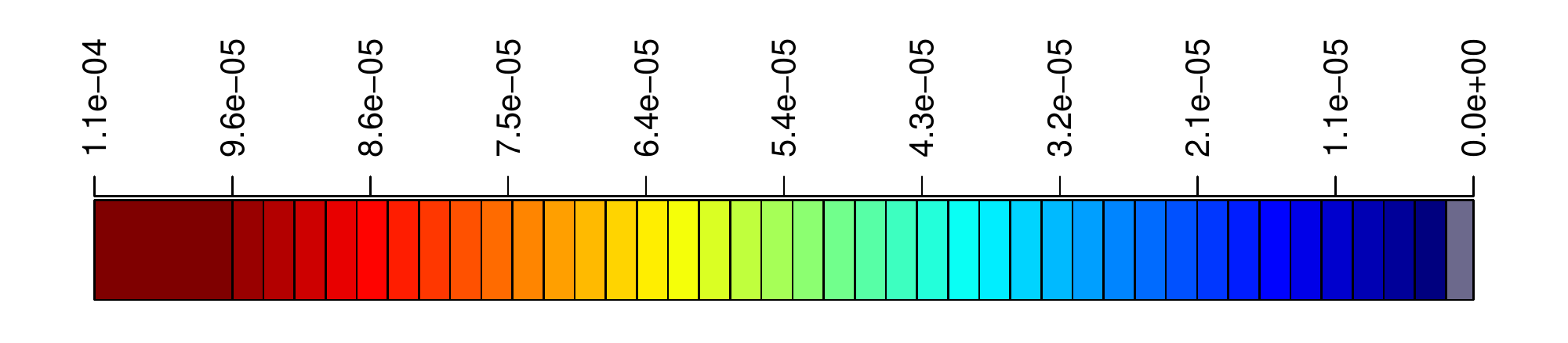} 
\hspace{-1.cm}
\includegraphics[width=\textwidth,angle=0]{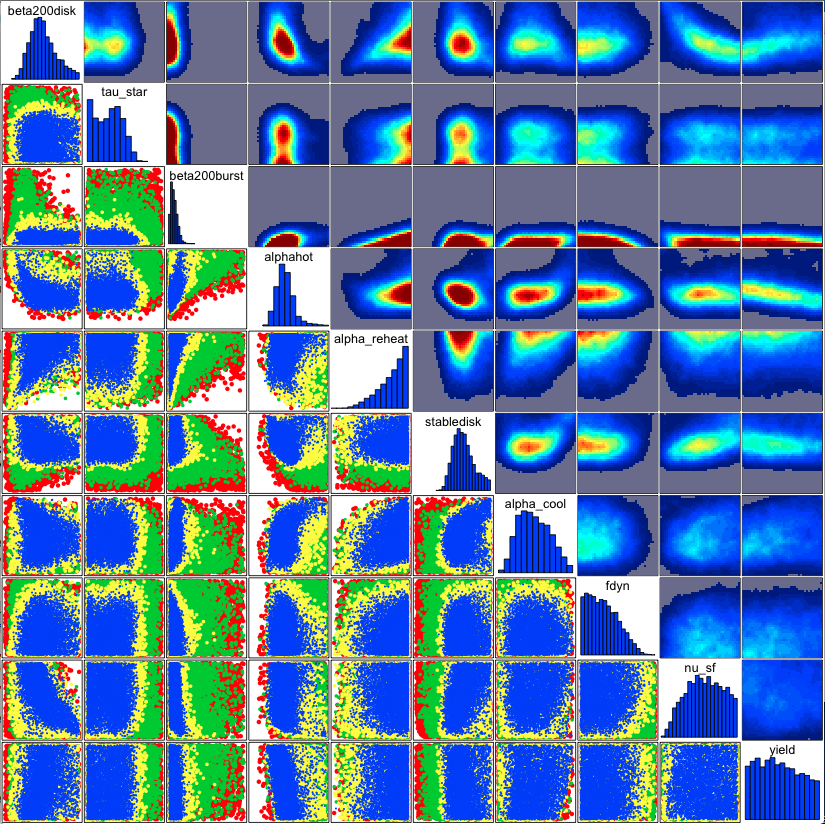}  
\caption{Results of the IDEMC algorithm when applied to the GALFORM model. The panels above the diagonal show the 2-dimensional optical depth plots, formed from the 5000 uniform samples over the wave 4 NROY space, for the 10 most active inputs (see table~\ref{tab_inputs}). These give the thickness or depth of the NROY space conditioned on two inputs, with the colours representing the depths as proportions of the input space, as given by the key above.  The main diagonal shows the corresponding 1-dimensional optical depth plots. Below the diagonal: the 5000 uniform samples from chromosomes 4, 6, 10 and 14, coloured red, green, yellow and blue respectively. These chromosomes were chosen as they each explore the NROY space after waves 1 to 4 respectively, and hence summarise the progression of the history match analysis.}\label{fig_galf_impdep}
\end{center}
\efi


As can be seen from figure~\ref{fig_galf_impdep}, the history match led to many active inputs being found, with several scientifically interesting interactions that combine to create a challenging shape for exploration of the NROY space.
Note that the size and shape of the NROY space would mean it would be utterly impracticable to explore this region without the use of fast emulators, and specifically emulators that have been trained to be highly accurate in the localised region around NROY space, through use of the iterative wave procedure. 
In many cases, inputs that are involved in modelling the same physical process show clear interrelations e.g. the inputs $\beta_{\rm 200,disk}$, $\beta_{\rm 200,burst}$, $\alphahot$ and $\alphareheat$ are all related to Supernova Feedback (see table~\ref{tab_inputs}), and this process is clearly critical in the accurate modelling of the stellar mass function. 
Other cross process dependencies can be seen such as the $\alphahot$, $\yield$ relation and the $\beta_{\rm 200,disk}$, $\nu_{sf}$ relation, suggesting clear links between the Supernova Feedback and both the Metal Enrichment and Star Formation processes respectively. 
The input $\tau_{\rm star}$ also appears to have a clear bi-modality (see for example the $\tau_{\rm star}$, $\alphahot$ plot), which could lead in future waves to the NROY space being composed of two disconnected pieces: again this should not pose a problem for the IDEMC algorithm, as is discussed and demonstrated in sections~\ref{sec:3d} and \ref{sec:10d}.

Often such information, in addition to the tradeoffs between inputs required to remain within NROY space, is of most interest to the cosmologists~\citep{Rodrigues13}, and can be readily obtained from a history match combined with the IDEMC algorithm. This avoids having to go through a possibly unwarranted full Bayesian calibration, which would involve the specification of high-dimensional joint distributions on all relevant uncertain quantities, often along with several further assumptions.

The plots below the diagonal in figure~\ref{fig_galf_impdep} show the 5000 uniform sample points from the IDEMC algorithm for chromosomes 4, 6, 10 and 14, coloured red, green, yellow and blue respectively.
These chromosomes explore the NROY space as defined after waves 1, 2, 3 and 4 and hence show the progression of the history match toward the final NROY space, defined at wave 4. 
In this way, the IDEMC algorithm output as represented by the complete set of uniform draws for all chromosomes
gives a comprehensive summary of the full history match. The higher chromosomes can be used to analyse which parts of the space are cut out progressively at each wave, and also which outputs are the most influential in contributing to such space reductions which is often of major importance when considering improvements to the GALFORM model, or indeed when planning more accurate future observations~\citep{McNeall13}.

At this stage of the project we are mainly interested in the NROY space defined using the implausibility cuts given in table~\ref{tab:waves}. However, as discussed in section~\ref{al.efficiency} it may be of interest to run the IDEMC algorithm to explore lower values of the implausibility function, e.g $\mathcal{I}_{M}< 2$ say. This would be sensible provided the emulators' accuracy is high enough to give meaning to small implausibility values and provided the outputs chosen for emulation are reasonably sufficient for the full set of 29 outputs. Though we don't believe this is the case at this wave of our analysis with GALFORM, we used IDEMC to discover whether, as currently defined, there are any settings of the model parameters with $\mathcal{I}_{M}< 2$. \par Despite millions of random samples failing to find a single point with $\mathcal{I}_{M}< 2$, the IDEMC algorithm discovered a region of parameter space that was roughly $3\times10^{-16}$ times the volume of the original space and was able to generate 50,000 uniform random samples from it. We do not show the results of a sample from this space as we don't believe that the emulator accuracy is high enough to make this an analysis of interest to the cosmologists. However, we include this description to illustrate that the need to be able to explore such tiny subregions of parameter spaces can arise in serious applications. We note that the size of this space is such that rejection sampling is infeasible, but even if the computing power existed to make it feasible, using the calculations outlined in section \ref{al.efficiency}, IDEMC is over $275$ \textit{billion} times more efficient based on generating $50,000$ uniform samples. 

As described in section~\ref{ap.to.gal}, the IDEMC algorithm is far more efficient that previous approaches, and this relative efficiency will only improve were we to perform subsequent waves, as the NROY space would become smaller still.
However, the next major stage in the GALFORM project is to introduce new output data sets into the history match, the first of these being the specific star formation rates~\citep{Rodrigues13}. These new outputs have not previously been matched by this version of the GALFORM model and we would expect that their inclusion would result in NROY space becoming significantly smaller at the next wave and possibly even zero in size. Use of the IDEMC algorithm will therefore become even more vital as this next stage of the scientific project proceeds.

\section{Discussion}\label{discussion}
We have presented the Implausibility Driven Evolutionary Monte Carlo algorithm (IDEMC) for obtaining uniform samples from subregions of computer model input spaces identified as worthy of further study by history matching. Though this method is developed in the context of history matching, it can be adapted to generate uniform samples from subspaces defined by a membership rule that consists of a continuous function passing a threshold test.  We have demonstrated the effectiveness of the algorithm in generating uniform samples over very small, possibly disconnected, subspaces. We showed that, for target spaces less than around $0.1\%$ of the volume of the original space, IDEMC is more efficient than the currently used rejection sampling methods, and showed that as this target volume is further reduced, the efficiency of our algorithm can be many orders of magnitude better than current methods. \par We applied IDEMC to a state of the art galaxy simulation model called GALFORM following a history matching exercise that had reduced the volume of the target space to around $10^{-5}$ of the volume of the original space. We used IDEMC to generate $5000$ uniform samples from the target space and discovered a number of interesting features of the not ruled out yet (NROY) space, including a potential bimodality in one of the parameters. \par The work with GALFORM will continue with the samples from this algorithm likely to be used to generate a 5th ensemble from GALFORM and a 5th wave of history matching, once additional datasets have been included in the analysis. Though we showed that IDEMC was only $18$ times more efficient than standard rejection sampling in our application, if a 5th wave of matching reduced NROY space by a similar amount as in previous waves, we would expect the target space to be around $10^{-7}$ times the volume of the original and our algorithm to be approximately $2000$ times more efficient than rejection sampling. In spaces this small, we believe rejection sampling is no longer feasible. \par
The question of how one should design runs for multi-wave computer experiments is an open, important and complex one with many possible considerations. We believe that the ability to efficiently generate uniform samples from the sorts of subspaces of interest located during these experiments offers a useful starting point from which to develop optimal designs with respect to whatever criteria are considered important. \par
For example, one possible design goal might be to ``fill'' NROY space by providing a design that covers the greatest amount of NROY space possible subject to any budget constraints. In order to meet this goal, the difficulty is in defining some measure of ``coverage'' with respect to the shape of NROY space. However, a very large uniform sample from NROY space is a good place to start, either in the development of such a measure or in selecting an $n$-point design from this sample that maximises coverage with respect to a previously defined measure. \par Optimal multi-wave designs need not be ``space filling'', and may depend on the ultimate goals of the experiment or the goal of the next wave. For example, in the case of history matching and when it is known that the next ensemble will be used to further reduce NROY space, an optimal design may be considered to be a design that minimizes the expected volume of NROY space. In cases where optimal design is considered to be an $n$-point design that maximises some criterion, IDEMC would be a natural and efficient tool for generating a large number of samples from which to select candidate designs. \par One of the great benefits of history matching  over Bayesian calibration is that the resulting NROY space might be empty. That is, the result of the exercise may be that there is no best input (a result that is not possible with calibration). However, supposing NROY space were empty, it would be extremely difficult to discover this using rejection sampling. Our 3rd illustrative example showed a space $10^{-18}$ times the volume of the original volume of the input space. It may not be feasible, in spaces of this volume to find any samples at all by rejection sampling in a reasonable amount of time. Hence, with rejection sampling, it may not be feasible to tell the difference between the result of a history matching being a tiny but important region of input space leading to excellent matches to real world observations and an empty NROY space indicating an inadequate computer model. \par IDEMC via the implausibility ladder selection algorithm of section 4 offers an efficient and targeted search for NROY space. If this space were empty, the selected implausibility levels will begin to converge on some number greater than the target, indicating that NROY space may be empty. Further work in this area will look to develop a stopping rule for the ladder selection algorithm based on the form of the emulators for those cases with NROY space empty.

\section{Acknowledgements}
The authors would like to thank R. G. Bower and L.F.S. Rodrigues for useful discussions and for providing the GALFORM data, and the  GALFORM group at the Institute for Computational Cosmology, Durham University, UK, for use of their resources. Ian Vernon is in part funded by an MRC grant. Daniel Williamson was funded by NERC RAPID-RAPIT project (NE/G015368/1) when completing this work.

\newcommand{\enquote}[1]{``#1''}

\section*{Appendix}\label{appendix.a}
In this appendix we present mixing plots for the various IDEMC algorithms applied in this paper.
\begin{figure}
\begin{center}
\hspace{-0.5cm}
\includegraphics[scale=0.63,angle=0]{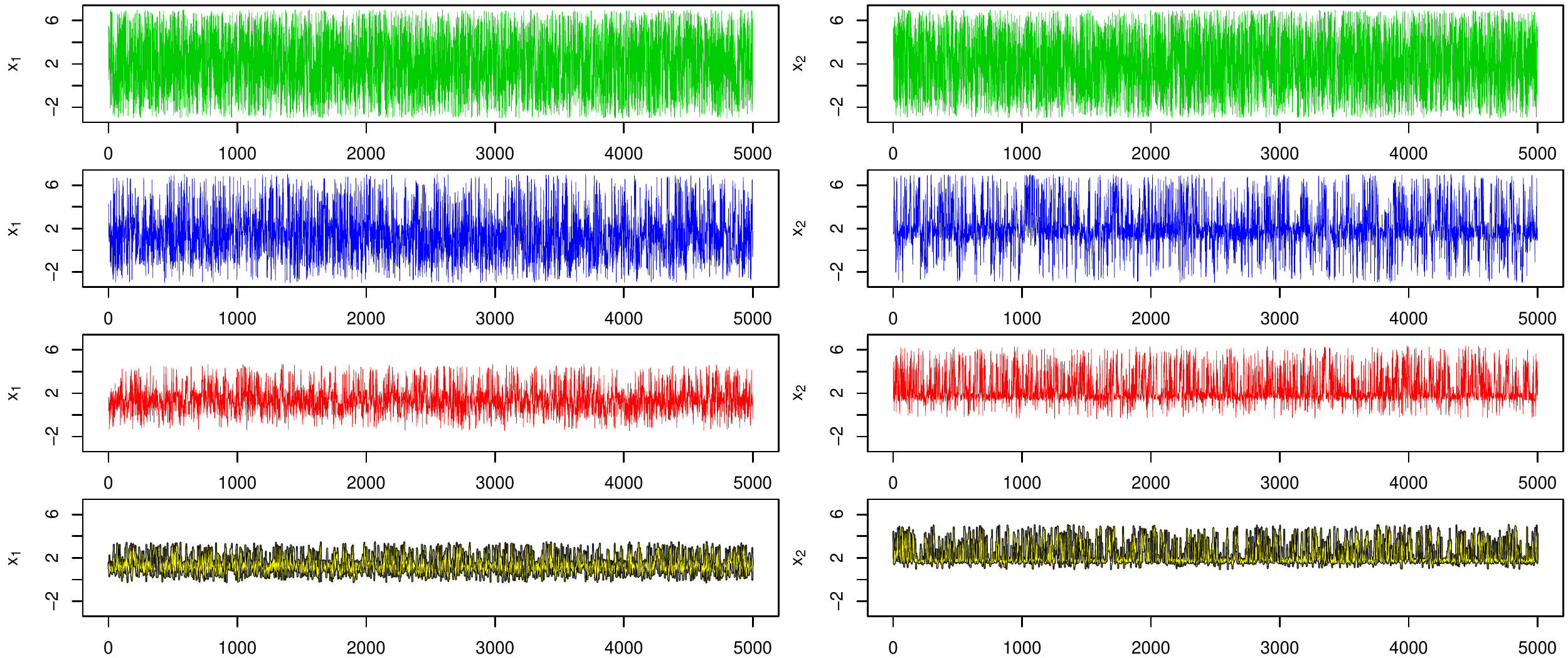}  
\end{center}
\caption{Mixing plots for the simple 2-dimensional example from section~\ref{ssec_2d}, for all four chromosomes (top to bottom), for inputs $x_1$ and $x_2$ (left column, right column respectively). The colours are consistent with those of figure~\ref{fig_2d}, right panel.
}\label{fig_2d_mixing}
\end{figure}

\begin{figure}
\begin{center}
\hspace{-0.6cm} \includegraphics[scale=0.54,angle=0]{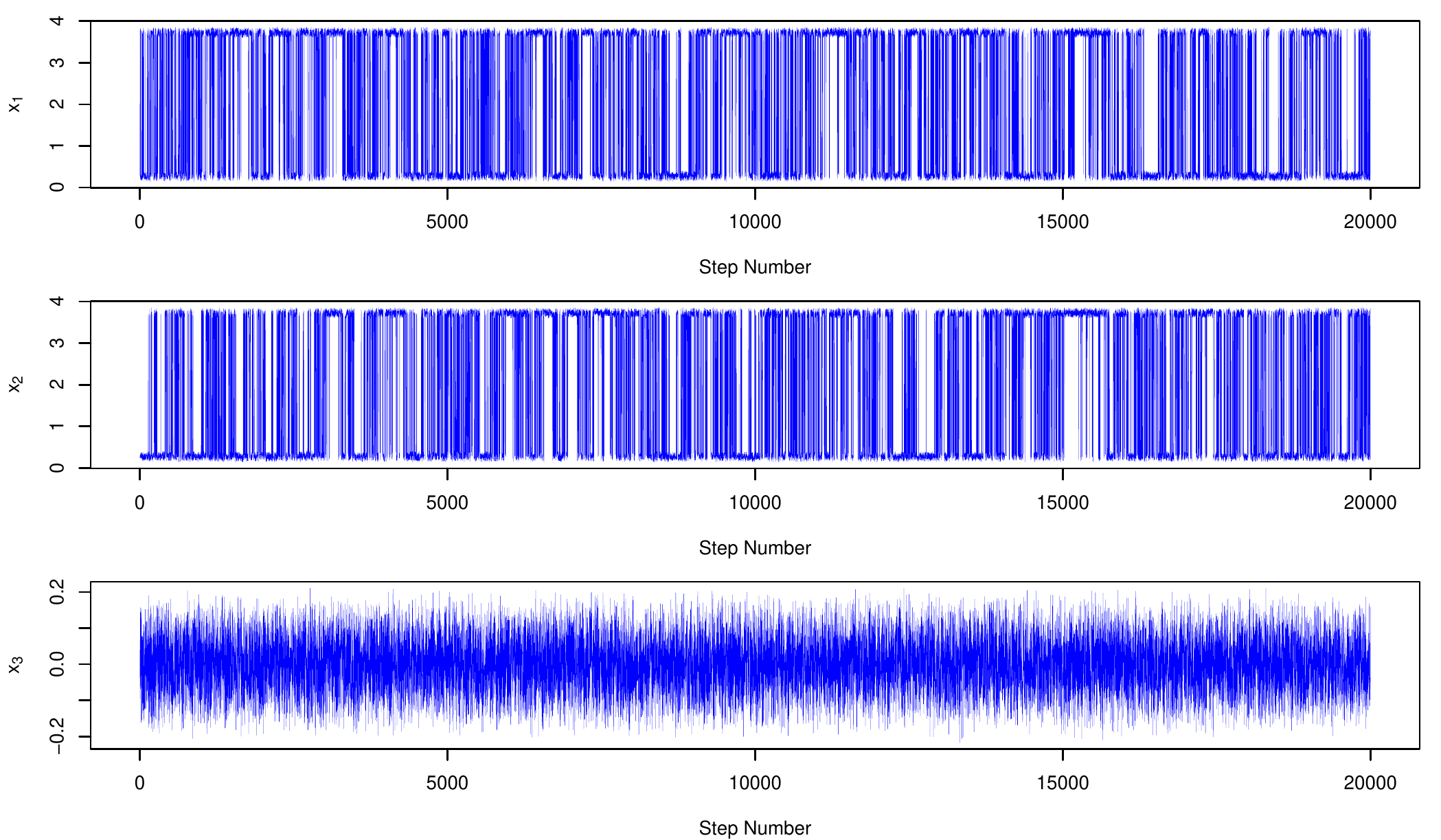}  
\end{center}
\caption{Mixing plots for the 3-dimensional 4 mode example. Note the excellent mixing in all three dimensions. 
}\label{fig_3d_mixing}
\end{figure}

\begin{figure}
\centering
\subfigure[][]{\includegraphics[height=0.33\textheight,width=0.4\textwidth]{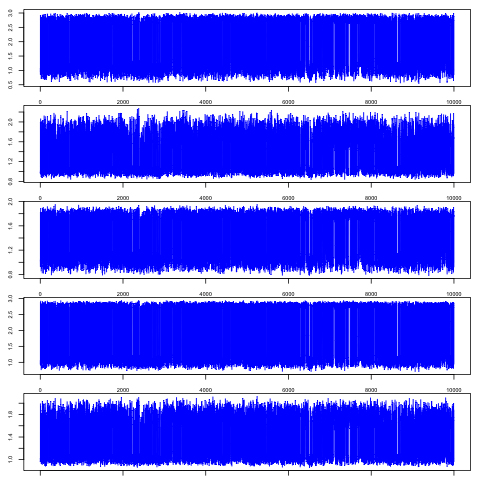}}
\subfigure[][]{\includegraphics[height=0.33\textheight,width=0.4\textwidth]{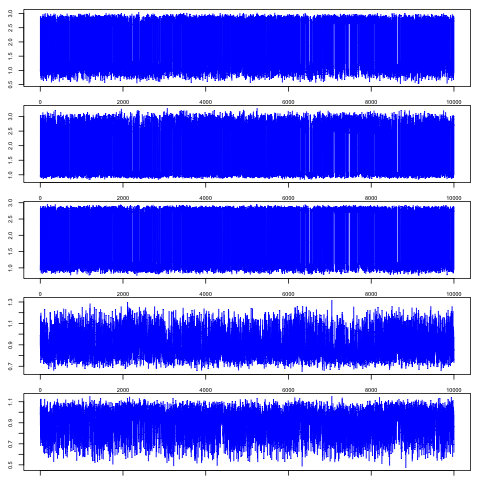}}
\caption{Trace plot for thinned target chromosome of the 10 dimensional example, again exhibiting excellent mixing.}\label{trace10d}
\end{figure}

\bfi
\begin{center}
\hspace{-1.5cm}
\includegraphics[scale=0.41,angle=0]{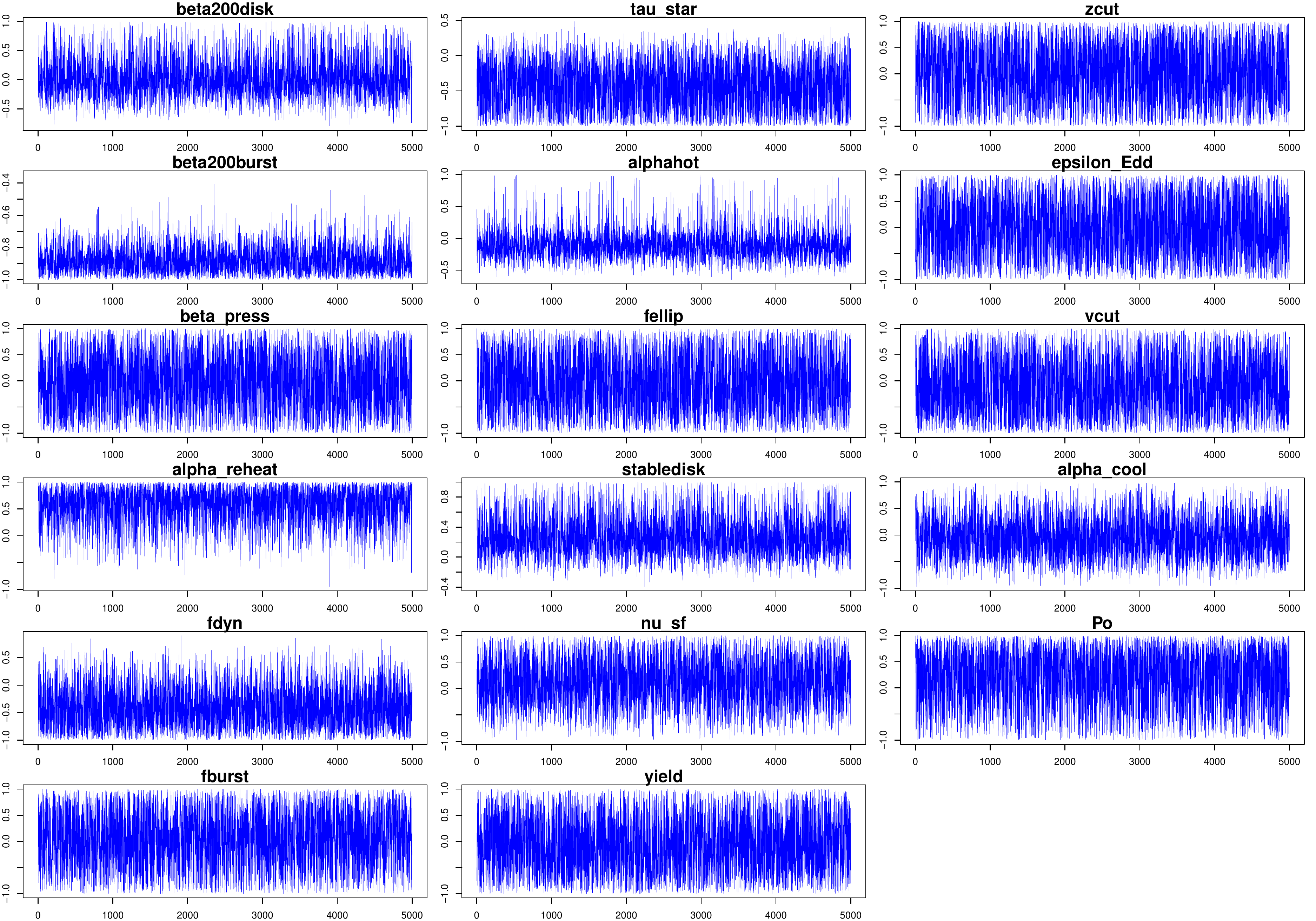}  \hspace{-1.0cm}
\caption{Trace plot for the thinned target chromosome of the GALFORM example, displaying good mixing. This is the 14th chromosome which explores the target wave 4 NROY space, as shown below the diagonal in figure~\ref{fig_galf_impdep} as the blue points.
}\label{fig_galf_mixing}
\end{center}
\efi

\end{document}